\documentclass[12pt]{article}
\usepackage{epsfig}
\usepackage{amsmath}
\usepackage{amsfonts}

\def\psl{\hbox{/\kern-.5800em$p$}}
\def\gappeq{\mathrel{\rlap {\raise.5ex\hbox{$>$}}
{\lower.5ex\hbox{$\sim$}}}}
\def\lappeq{\mathrel{\rlap{\raise.5ex\hbox{$<$}}
{\lower.5ex\hbox{$\sim$}}}}

\begin{document}
\pagestyle{empty}
\begin{flushright}
UMN-TH-2429/06\\
\end{flushright}
\vspace*{5mm}

\begin{center}
{\Large\bf Les Houches Lectures on\\
\vspace{0.3cm}
Warped Models and Holography\footnote{Based on lectures given 
at the Les Houches Summer School - Session LXXXIV, {\it Particle Physics
Beyond the Standard Model}, August 1-26, 2005.}}
\vspace{1.0cm}

{\sc Tony Gherghetta}\\
\vspace{.5cm}
{\it\small {School of Physics and Astronomy\\
University of Minnesota\\
Minneapolis, MN 55455, USA}}\\
\vspace{.5cm}
{\small\tt tgher@physics.umn.edu}
\end{center}

\vspace{1cm}
\begin{abstract}

The theoretical tools required to construct models in warped extra dimensions 
are presented. This includes how to localise zero modes in the warped bulk 
and how to obtain the holographic interpretation using the AdS/CFT 
correspondence. Several models formulated in warped space are then 
discussed including nonsupersymmetric and supersymmetric theories as well 
as their dual interpretation. Finally it is shown how grand unification occurs
in warped models.

\end{abstract}

\vfill
\begin{flushleft}
\end{flushleft}
\eject
\pagestyle{empty}
\setcounter{page}{1}
\setcounter{footnote}{0}
\pagestyle{plain}

\tableofcontents

\newpage
\section{Introduction}

Warped extra dimensions have provided a new framework for addressing the
hierarchy problem in extensions of the Standard Model. The curved fifth 
dimension is compactified on a line segment where distance scales 
are measured with the nonfactorisable metric of anti-de Sitter (AdS) space. 
Since distance and hence energy scales are location dependent in AdS
space, the hierarchy problem can be redshifted away. However this novel 
solution also has a more mundane explanation. By the AdS/CFT correspondence, 
five-dimensional (5D) AdS space has a dual four-dimensional (4D) 
interpretation in terms of a strongly-coupled 
conformal field theory (CFT). In this 4D guise the hierarchy problem 
is solved by having a low cutoff scale that is associated with the 
conformal symmetry breaking scale of the CFT. Both interpretations are 
equally valid and allow the infrared (IR) scale to be hierarchically smaller 
and stable compared to the ultraviolet (UV) scale. This is much like the 
photon and electron 
of QED compared with the baryons and mesons of QCD. The Planck scale, $M_P$ 
provides the cutoff scale for QED, while in QCD the composite baryons and 
mesons are only valid up to the QCD scale $\Lambda_{QCD} \ll M_P$. In 
warped models of electroweak physics the Higgs is composite at the TeV scale, 
while the Standard Model fermions and gauge bosons are partly composite to 
varying degrees, ranging from an elementary electron to a composite top quark. 
The Planck scale again provides the UV cutoff for the elementary states,
while the TeV scale is the compositeness or IR scale.

This hybrid framework of electroweak physics has led to a renaissance of 
composite Higgs models that have many desirable features. They are 
consistent with electroweak precision data that place strong bounds 
on the scales of new physics. New physics at the TeV scale generically 
leads to flavor problems, but these are absent in warped models, 
since a GIM-like mechanism operates. The warped models also explain the 
fermion mass hierarchy (including neutrino masses) as well as incorporate 
grand unification with logarithmic running and unified gauge couplings. 
This sophisticated level of model building is due to the complementary 
nature of the AdS/CFT framework. The 5D warped bulk not only provides a 
simple geometric picture in which to construct these models, but more 
importantly is also a weakly-coupled description in which calculations can be 
performed. In fact this theoretical tool provides a unique window into 
strongly-coupled 4D gauge 
theories and goes beyond the application to electroweak physics that will 
be the primary concern of these lectures. The 5D picture is complemented by 
the 4D description which is more intuitive, primarily from our understanding 
of QCD, except that the 4D gauge theory is strongly-coupled and a 
perturbative analysis is not possible.

Hence the plan of these lectures will be to begin in 5D warped space 
and describe how to localise zero modes anywhere in a slice of AdS. 
This will enable us to quite simply construct the Standard Model in the bulk. 
The more intuitive dual 4D interpretation will then be obtained by writing 
down the AdS/CFT dictionary. This dictionary will allow us to give either a 
5D or 4D description for any type of warped model. The AdS/CFT framework
need not only apply to electroweak physics. Supersymmetry will be 
subsequently introduced and the model-building possibilities surveyed. 
Lastly the novel features of grand unification in warped space 
will be discussed.

Since the aim of these lectures is to emphasize the dual nature of warped 
models a little background knowledge of warped extra dimensions 
will be useful. This is nicely reviewed in Refs~\cite{vr, gg, cc,apl, 
raman}. Some theoretical aspects of warped models presented in these 
lectures have also been covered in Refs~\cite{cc,raman,chm}. Higgless
models are not covered in these lectures, but these type of warped models
are discussed in Refs~\cite{chm,cg}. Finally, while these lectures 
concentrate on the theoretical aspects of warped models, the phenomenological
aspects are just as important, and these can be found in Ref~\cite{chm,jh}.

\section{Bulk fields in a slice of AdS$_5$}

\subsection{A slice of AdS$_5$}\label{slice}

\begin{figure}[ht]
\centerline{ 
   \epsfxsize 2.5  truein \epsfbox {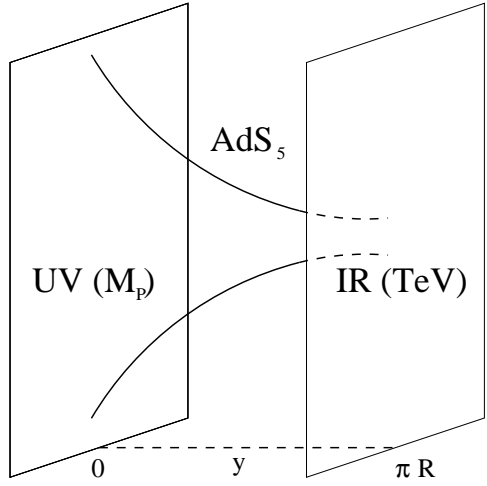}
    }

\caption{\it A slice of AdS$_5$: The Randall-Sundrum scenario.}
\label{sliceAdS}
\end{figure}

Let us begin by considering a 5D spacetime with the AdS$_5$ metric
\begin{equation}
\label{adsmetric}
    ds^2 = e^{-2ky} \eta_{\mu\nu} dx^\mu dx^\nu + dy^2 \equiv 
      g_{MN} dx^M dx^N,
\end{equation}
where $k$ is the AdS curvature scale. The spacetime indices 
$M=(\mu,5)$ where $\mu = 0,1,2,3$ and $\eta_{\mu\nu}={\rm diag}(-+++)$ is 
the Minkowski metric.
The fifth dimension $y$ is compactified on a $Z_2$ orbifold 
with a UV (IR) brane located at the orbifold fixed points $y^\ast=0 (\pi R)$.
Between these two three-branes the metric (\ref{adsmetric}) is a solution
to Einstein's equations provided the bulk cosmological constant and 
the brane tensions are appropriately tuned (see, for example, the lectures by 
Rubakov~\cite{vr}). This slice of AdS$_5$ is the 
Randall-Sundrum solution~\cite{rs1} (RS1) and is geometrically depicted in 
Fig.\ref{sliceAdS}.

In RS1 the Standard Model particle states are  
confined to the IR brane. The hierarchy problem is then solved by noting that 
generic mass scales $M$ in the 5D theory are scaled down to $M e^{-\pi kR}$ 
on the IR brane at $y=\pi R$. In particular since the Higgs boson $H$ is 
localised on the IR brane this means that the dimension two Higgs mass term 
gets rescaled by an amount 
\begin{equation}
        m_H^2 |H|^2 \rightarrow (m_H e^{-\pi kR})^2 |H|^2~,
\end{equation}
so that a Higgs mass parameter $m_H\sim {\cal O}(M_5)$ in the 5D theory 
will naturally be associated with a hierarchically smaller scale on 
the IR brane (where $M_5$ is the 5D fundamental mass scale). However on the 
IR brane higher-dimension operators with dimension greater than four, 
such as those associated with proton decay, flavour changing neutral 
currents (FCNC) and neutrino masses will now be suppressed by the warped 
down scale
\begin{eqnarray}
       \frac{1}{M_5^2} \bar\Psi_i\Psi_j\bar\Psi_k\Psi_l& \rightarrow&
        \frac{1}{(M_5 e^{-\pi kR})^2} \bar\Psi_i\Psi_j\bar\Psi_k\Psi_l ~,\\
        \frac{1}{M_5}\nu\nu H H &\rightarrow& \frac{1}{M_5 e^{-\pi k R}} 
    \nu\nu H H~,
\end{eqnarray}
where $\Psi_i$ is a Standard Model fermion and $\nu$ is the neutrino.
This leads to generic problems with proton decay and FCNC effects, 
and also neutrino masses are no longer consistent with experiment. 
Thus, while the hierarchy problem has been addressed in the Higgs sector 
by a classical rescaling of the Higgs field, this has come at the expense of 
introducing proton decay and FCNC problems from higher-dimension
operators that were sufficiently suppressed in the Standard Model. 
\\
\\
\noindent
$\bullet$ {\sc Exercise:} {\it The classical rescaling $\Phi 
\rightarrow e^{d_{\Phi} \pi k R} \Phi$ where $d_\Phi=1 (\frac{3}{2})$ 
for scalars (fermions), suffers from a quantum anomaly and leads to the 
addition of the Lagrangian term
\begin{equation}
   \delta {\cal L}_{\rm anomaly} = \pi kR\sum_i  
      \frac{\beta(g_i)}{4 g_i^3} {\rm Tr}\, F_{\mu\nu,i}^2~,
\end{equation}
where $\beta(g_i)$ is the $\beta$-function for the corresponding 
gauge couplings $g_i$. Show that this anomaly implies that quantum 
mass scales, such as the gauge coupling unification scale $M_{GUT}$, are 
also redshifted by an amount $M_{GUT} e^{-\pi k R}$.}
\\
\\
Instead in the slice of AdS$_5$ with the Standard Model fields confined on 
the IR brane one has to resort to discrete symmetries to forbid the 
offending higher-dimension operators. Of course 
it is not adequate just to forbid the leading higher-dimension operators. 
Since the cutoff scale on the IR brane is low (${\cal O}$(TeV)), 
successive higher-dimension operators must also be eliminated to very 
high order.

This feature of RS1 stems from the fact that all Standard Model particles 
are localised on the IR brane. However to address the hierarchy problem,
only the Higgs field needs to be localised on the IR brane. The Standard
Model fermions and gauge fields do not have a hierarchy problem 
and therefore can be placed anywhere in the 
bulk~\cite{chnoy,gp,dhr}.
In this way the UV brane can be used to provide a sufficiently high
scale to help suppress higher-dimension operators while still solving the
hierarchy problem~\cite{gp}.

\subsection{The bulk field Lagrangian}
Let us consider fermion $\Psi$, scalar $\Phi$ and vector $A_M$ bulk fields.
In a slice of AdS$_5$ the 5D action is given by
\begin{eqnarray}
\label{5daction}
     S_5 &=&-\int d^4 x~dy~\sqrt{-g}\left[ \frac{1}{2} M_5^3 R 
     + \Lambda_5 \right.\nonumber\\
     && \qquad\qquad\qquad+~\frac{1}{4 g_5^2} F_{MN}^2 + |D_M \Phi|^2 
    + i{\bar\Psi} \Gamma^M \nabla_M \Psi \nonumber\\
    &&\qquad\qquad\qquad+~\left. m_\phi^2 |\Phi|^2 
    + i m_\psi \bar\Psi\Psi \right]~,
\end{eqnarray}
where $M_5$ is the 5D fundamental scale, $\Lambda_5$ is the 
bulk cosmological constant and $g_5$ is the 5D gauge coupling.
In curved space the gamma matrices are $\Gamma_M = e_M^A\gamma_A$,
where $e_M^A$ is the funfbein defined by $g_{MN}= e_M^A e_N^B \eta_{AB}$ 
and $\gamma_A=(\gamma_\alpha,\gamma_5)$ are the usual gamma matrices in flat 
space. The curved space covariant derivative $\nabla_M =D_M+\omega_M$,
where $\omega_M$ is the spin connection and $D_M$ is the gauge
covariant derivative for fermion and/or scalar fields charged under some 
gauge symmetry. The action (\ref{5daction}) includes all terms to quadratic 
order that are consistent with gauge symmetries and general coordinate 
invariance. In particular this only allows a mass term $m_\phi$ for the bulk 
scalar and a mass term $m_\psi$ for the bulk fermion.

In general the equation of motion for the bulk fields is obtained by
requiring that $\delta S_5 =0$. This variation of the action (\ref{5daction})
can be written in the form 
\begin{equation}
\label{var5daction}
      \delta S_5 = \int  d^5 x~\delta\phi~({\cal D}\phi)
        +\int  d^4 x~ \delta\phi~({\cal B}\phi) \big |_{y^\ast}~,
\end{equation}
where $\phi$ is any bulk field. Requiring the first term in (\ref{var5daction})
to vanish gives the equation of motion ${\cal D}\phi=0$. However the second 
term in (\ref{var5daction}) is evaluated at the boundaries $y^\ast$ of the 
fifth dimension $y$. The vanishing of the second term thus leads to the 
boundary conditions $\delta\phi|_{y^\ast}=0$ or ${\cal B}\phi|_{y^\ast}=0$. 
Note that there 
are also boundary terms arising from the orthogonal directions $x^\mu$, but 
these are automatically zero because $\phi$ is assumed to vanish at the 
4D boundary $x^\mu =\pm \infty$.

\subsubsection{Scalar fields}
Suppose that the bulk scalar field has a mass squared 
$m_\phi^2= a k^2$ where we have defined the bulk scalar mass in units of 
the curvature scale $k$ with dimensionless coefficient $a$. The equation 
of motion derived from the scalar part of the variation of the 
action (\ref{5daction}) is
\begin{equation}
        \partial^2 \Phi + e^{2 ky}\partial_5 (e^{-4k y} \partial_5 \Phi) 
    -m_\phi^2 e^{-2 ky}\Phi = 0,
\end{equation}
where $\partial^2 = \eta^{\mu\nu}\partial_\mu\partial_\nu$. We are interested 
in the zero mode solution of this equation. The solution is obtained
by assuming a separation of variables 
\begin{equation}
     \Phi(x,y)=\frac{1}{\sqrt{\pi R}} \sum_n \Phi^{(n)}(x) \phi^{(n)}(y)~,
\end{equation}
where $\Phi^{(n)}$ are the Kaluza-Klein modes satisfying 
$\partial^2 \Phi^{(n)}(x)= m_n^2 \Phi^{(n)}(x)$ and $\phi^{(n)}(y)$ is the 
profile of the Kaluza-Klein mode in the bulk. The general solution for the zero
mode $(m_0=0)$ is given by
\begin{equation}
    \phi^{(0)}(y) = c_1~e^{(2-\alpha)k y}+c_2~e^{(2+\alpha)k y}~,
\end{equation}
where $\alpha\equiv\sqrt{4+a}$ and $c_1,c_2$ are arbitrary constants. 
These constants can be determined by imposing boundary conditions at the 
brane locations, which following from the second term of (\ref{var5daction}) 
can be either Neumann $\partial_5 \phi^{(n)} |_{y^\ast} = 0$ or 
Dirichlet $\phi^{(n)}|_{y^\ast} =0$. However for $a\neq 0$ imposing Neumann 
conditions leads to $c_1=c_2=0$. Similarly Dirichlet conditions lead to 
$c_1=c_2=0$. This implies that there is no zero mode solution with
simple Neumann or Dirichlet boundary conditions.

Instead in order to obtain a zero mode we need to modify the boundary 
action and include boundary mass terms~\cite{gp}
\begin{equation}
\label{sbdy}
     S_{bdy}= -\int d^4 x~dy~\sqrt{-g}~2~b~k \left[ \delta(y) - 
     \delta(y-\pi R) \right] |\Phi|^2~,
\end{equation}
where $b$ is a dimensionless constant parametrising the boundary mass in 
units of $k$. The Neumann boundary conditions are now modified to
\begin{equation}
\label{scalarbc}
  \left(\partial_5 \phi^{(0)} - b~k~\phi^{(0)}\right)\bigg|_{0,\pi R}=0.
\end{equation}

\noindent
$\bullet$ {\sc Exercise:} {\it Verify that the boundary conditions 
(\ref{scalarbc}) follow from (\ref{var5daction}) after including the 
boundary mass terms (\ref{sbdy}).}
\\
\\
Imposing the modified Neumann boundary conditions at $y^\ast =0,\pi R$ 
leads to the equations
\begin{eqnarray}
      (2-\alpha-b)~c_1 + (2+\alpha-b)~c_2& = &0~,\\
      (2-\alpha-b)~c_1~e^{(2-\alpha)\pi k R} + 
      (2+\alpha-b)~c_2~e^{(2+\alpha)\pi k R} &=& 0~.
\end{eqnarray}
These equations depend on the two arbitrary mass parameters $a$ and $b$. For 
generic values of these parameters the solution to the boundary conditions 
again leads to $c_1=c_2=0$. However, if $b=2-\alpha$ then this implies that 
only $c_2=0$, while if $b=2+\alpha$ then we obtain $c_1=0$. Note that in 
principle there are three mass parameters if we introduce two parameters 
corresponding to each boundary. However in (\ref{sbdy}) we have chosen the 
mass parameters on the two boundaries to be equal and opposite. Thus a 
nonzero part of the general solution always survives and the zero mode 
solution becomes
\begin{equation}
        \phi^{(0)}(y) \propto e^{b k y}~,
\end{equation}
where $b=2\pm \alpha$. Assuming $\alpha$ to be real (which requires 
$a\geq -4$ in accord with the Breitenlohner-Freedman bound~\cite{bf} for the 
stability of AdS space), the parameter $b$ has a range 
$-\infty < b < \infty$. The localisation features of the zero mode follows 
from considering the kinetic term
\begin{eqnarray}
    &&-\int d^5 x~\sqrt{-g}~g^{\mu\nu}~\partial_\mu \Phi^\ast \partial_\nu 
    \Phi + \dots \nonumber\\
&& =-\int d^5 x~e^{2(b-1) k y}~\eta^{\mu\nu}~\partial_\mu \Phi^{(0)\ast}(x)
    \partial_\nu \Phi^{(0)}(x)+\dots
\end{eqnarray}
Hence, with respect to the 5D flat metric the zero mode profile is given 
by
\begin{equation}
     \widetilde\phi^{(0)}(y) \propto e^{(b-1) k y} = e^{(1\pm\sqrt{4+a})ky}~.
\end{equation}
We see that for $b<1~(b >1)$ the zero mode is localised towards the UV (IR) 
brane and when $b=1$ the zero mode is flat. Therefore using the one remaining 
free parameter $b$ the scalar zero mode can be localised anywhere in the bulk.

The general solution of the  Kaluza-Klein modes corresponding to $m_n\neq 0$ 
is given by
\begin{equation}
    \phi^{(n)}(y) = e^{2 k y}\left[ c_1 J_\alpha \left(\frac{m_n}{k e^{-k y}}
    \right) + c_2 Y_\alpha \left(\frac{m_n}{k e^{-k y}}\right) \right]~,
\end{equation}
where $c_{1,2}$ are arbitrary constants. The Kaluza-Klein masses are 
determined by imposing the boundary conditions 
and in the limit $\pi kR \gg1$ lead to the approximate values
\begin{equation}
\label{scKK}
    m_n\approx \left(n+\frac{1}{2}\sqrt{4+a}-\frac{3}{4}\right)\pi 
     k~e^{-\pi k R}~.
\end{equation}
The fact that the Kaluza-Klein mass scale is associated with the IR scale 
($k e^{-\pi kR}$) is consistent with the fact that the Kaluza-Klein modes 
are localised near the IR brane, and unlike the zero mode can not be 
arbitrarily localised in the bulk.

\subsubsection{Fermions}
Let us next consider bulk fermions in a slice of AdS$_5$~\cite{gn,gp}. 
In five dimensions a fundmental spinor representation
has four components, so fermions are described by Dirac spinors $\Psi$. 
Under the $ Z_2$ symmetry $y\rightarrow -y$ a fermion transforms 
(up to a phase $\pm$) as 
\begin{equation}
\label{fermpar}
   \Psi(-y) = \pm\gamma_5 \Psi(y)~,
\end{equation}
so that $\bar\Psi \Psi$ is odd. 
Since only invariant (or even)  terms under the $Z_2$ symmetry can be added 
to the bulk Lagrangian the corresponding mass parameter for a fermion must 
necessarily be odd and given by
\begin{equation}
\label{bulkfm}
    m_\psi = c~k~(\epsilon(y) -\epsilon(y-\pi R))~,
\end{equation}
where $c$ is a dimensionless mass parameter and $\epsilon(y) = y/|y|$. 
We have again chosen the mass parameter $c$ to be equal and opposite on the
two boundaries. The corresponding equation of motion for fermions resulting 
from the action (\ref{5daction}) is
\begin{eqnarray}
\label{feom}
   e^{ky}\eta^{\mu\nu}\gamma_\mu\partial_\nu \widehat\Psi_- +
   \partial_5\widehat\Psi_+ + m_\psi\widehat\Psi_+ &=& 0~,\nonumber\\
   e^{ky}\eta^{\mu\nu}\gamma_\mu\partial_\nu \widehat\Psi_+ -
   \partial_5\widehat\Psi_- + m_\psi\widehat\Psi_- &=& 0~,
\end{eqnarray}
where $\widehat\Psi = e^{-2k y}\Psi$ and $\Psi_\pm$ are the components of 
the Dirac spinor $\Psi= \Psi_+ + \Psi_-$ with $\Psi_{\pm}=\pm\gamma_5 
\Psi_{\pm}$. Note that the equation of motion is now a first order coupled 
equation between the components of the Dirac spinor $\Psi$.
\\
\\
\noindent
$\bullet$ {\sc Exercise:} {\it Using the fact that the spin connection
for the AdS$_5$ metric is given by
\begin{equation}
    \omega_M = \left( \frac{k}{2}\gamma_5\gamma_\mu e^{-ky},~0\right)~,
\end{equation}
derive the bulk fermion equation of motion (\ref{feom}) from the
action (\ref{5daction}).}
\\
\\
The solutions of the bulk fermion equation of motion (\ref{feom})
are again obtained by separating the variables 
\begin{equation}
     \Psi_{\pm}(x,y)=\frac{1}{\sqrt{\pi R}} \sum_n \Psi_{\pm}^{(n)}(x) 
     \psi_{\pm}^{(n)}(y)~,
\end{equation}
where 
$\Psi_{\pm}^{(n)}$ are the Kaluza-Klein modes satisfying 
$\eta^{\mu\nu}\gamma_\mu\partial_\nu \Psi_{\pm}^{(n)} = m_n \Psi_{\pm}^{(n)}$.
The zero mode solutions can be obtained for $m_n=0$ and 
the general solution of (\ref{feom}) is given by
\begin{equation}
     \widehat\psi_{\pm}^{(0)}(y) = d_{\pm}~e^{\mp c ky}~,
\end{equation}
where $d_{\pm}$ are arbitrary constants. The $Z_2$ symmetry 
implies that one of the components $\psi_{\pm}$ must always be odd. If
$\gamma_5={\rm diag}(1,-1)$, then (\ref{fermpar}) implies that $\psi_\mp$
is odd and there is no corresponding zero mode for this component of $\Psi$. 
In fact this is how 4D chirality is recovered from the vectorlike 5D bulk 
and is the result of compactifying on the $Z_2$ orbifold. For the remaining
zero mode the boundary condition obtained from (\ref{var5daction}) with the
boundary mass term (\ref{bulkfm}) is the modified Neumann condition
\begin{equation}
     \left(\partial_5 \widehat\psi_\pm^{(0)} \pm 
    c~k~\widehat\psi_\pm^{(0)}\right)\bigg|_{0,\pi R} = 0~.
\end{equation}
Thus there will always be a zero mode since the boundary condition is 
trivially the same as the equation of motion. For concreteness let us choose
$\psi_-$ to be odd, then the only nonvanishing zero mode component of $\Psi$ is
\begin{equation}
     \psi_+^{(0)}(y) \propto e^{(2-c)ky}~.
\end{equation}
Again the localisation features of this mode are obtained by
considering the kinetic term
\begin{eqnarray}
    &&-\int d^5 x~\sqrt{-g}~g^{\mu\nu}~i\bar\Psi \Gamma_\mu \partial_\nu 
   \Psi + \dots \nonumber\\
&& = -\int d^5 x~e^{2(\frac{1}{2}-c) k y}~\eta^{\mu\nu}~i\bar\Psi_+^{(0)}(x)
    \gamma_\mu \partial_\nu \Psi_+^{(0)}(x)+\dots~.
\end{eqnarray}
Hence with respect to the 5D flat metric the fermion zero mode profile is
\begin{equation}
     \widetilde\psi_+^{(0)}(y) \propto e^{(\frac{1}{2}-c)ky}~.
\end{equation}
When $c> 1/2~(c<1/2)$ the fermion zero mode is localised towards
the UV (IR) brane while the zero mode fermion is flat for $c=1/2$.
So, just like the scalar field zero mode the fermion zero mode
can be localised anywhere in the 5D bulk. 

The nonzero Kaluza-Klein fermion modes can be obtained by solving the
coupled equations of motion for the Dirac components $\psi_{\pm}^{(n)}$.
This leads to a pair of decoupled second order equations that can
be easily solved. The expressions for the corresponding wave functions and 
Kaluza-Klein masses are summarised in the next subsection.

\subsubsection{Summary}
A similar analysis can also be done for the bulk graviton~\cite{rs2} and 
bulk gauge field~\cite{gf1,gf2}. In these cases there are no bulk or boundary 
masses and the corresponding graviton zero mode is localised on the UV brane
while the gauge field zero mode is flat and not localised in the 5D 
bulk~\footnote{Note however that bulk and boundary masses can be 
introduced for the graviton~\cite{gpp} and gauge field~\cite{bg}, thereby 
changing their localization profile and corresponding operator dimensions. 
But this is beyond the applications that will be considered in these 
lectures.}. A summary of the bulk zero mode profiles is given in Table 1.

\begin{table}[t]
\caption{\it The zero mode profiles of bulk fields and the corresponding CFT 
operator dimensions.}
\label{tab1}
\tabcolsep=0.1cm
\begin{tabular*}{\columnwidth}{@{\extracolsep{\fill}}@{}ccc@{}}
\hline
Field  & Profile & dim $\cal O$ \\
\hline
scalar $\phi^{(0)}(y)$ &  $e^{(1\pm\sqrt{4+a})ky}$ & $2+\sqrt{4+a}$\\
fermion  $\psi^{(0)}_+(y)$ & $e^{\left(\frac{1}{2}-c\right)ky}$ & 
$\frac{3}{2}+|c+ \frac{1}{2}|$ \\
vector $A_\mu^{(0)}(y)$ & 1 & 3\\
graviton $ h_{\mu\nu}^{(0)}(y)$ & $e^{-ky}$ & 4\\
\hline
\end{tabular*}
\end{table}

Similarly the Kaluza-Klein mode ($m_n\neq 0$) solutions can be obtained for 
all types of bulk fields and combined into one general expression~\cite{gp}
\begin{equation}
\label{genwf}
    f^{(n)}(y) = \frac{e^{\frac{s}{2} k y}}{N_n}\left[ J_\alpha 
    \left(\frac{m_n}{k e^{-k y}}\right) + b_\alpha Y_\alpha \left(\frac{m_n}
     {k e^{-k y}}\right) \right]~,
\end{equation}
for $f^{(n)}=(\phi^{(n)},\widehat\psi_{\pm}^{(n)},A_\mu^{(n)})$
where 
\begin{equation}
\label{balpha}
    b_\alpha = -\frac{(-r+\frac{s}{2})J_\alpha(\frac{m_n}{k})+\frac{m_n}{k}
     J'_\alpha(\frac{m_n}{k})}{(-r+\frac{s}{2})Y_\alpha(\frac{m_n}{k})
     +\frac{m_n}{k}Y'_\alpha(\frac{m_n}{k})}~,
\end{equation}
and 
\begin{equation}
     N_n \simeq \frac{1}{\sqrt{\pi^2 R~m_n~e^{-\pi kR}}}~,
\end{equation}
with $s=(4,1,2)$, $r=(b,\mp c,0)$ and 
$\alpha=(\sqrt{4+a}, |c\pm \frac{1}{2}|, 1)$. The graviton modes 
$h_{\mu\nu}^{(n)}$ are identical to the scalar modes $\phi^{(n)}$ except 
that $a=b=0$. The Kaluza-Klein mass spectrum is approximately given by
\begin{equation}
\label{KKspect}
    m_n\simeq \left(n+\frac{1}{2}(\alpha-1)\mp\frac{1}{4}\right)\pi 
     k~e^{-\pi kR}~,
\end{equation}
for even (odd) modes and $n=1,2,\dots$.
Note that the Kaluza-Klein modes for all types of bulk fields are always 
localised near the IR brane. Unlike the zero mode there is no freedom to 
delocalise the Kaluza-Klein (nonzero) modes away from the IR brane.

\section{The Standard Model in the Bulk}

We can now use the freedom to localise scalar and fermion zero mode 
fields anywhere in the warped bulk to construct a bulk Standard Model. 
Recall that the hierarchy problem only affects the Higgs boson.
Hence to solve the hierarchy problem the Higgs scalar field must
be localised very near the TeV brane, and for simplicity we will assume that
the Higgs is confined on the TeV brane (as in RS1). However we will
now consider the possible effects of allowing fermions and gauge bosons 
to live in the warped bulk.

\subsection{Yukawa couplings}

One consequence of allowing fermions to be localised anywhere in the bulk, 
is that Yukawa coupling hierarchies are naturally generated by separating 
the fermions from the Higgs on the TeV brane.
Each Standard Model fermion is identified with the zero mode 
of a corresponding 5D Dirac spinor $\Psi$. For example, the left-handed 
electron doublet $e_L$ is identified with the zero mode of $\Psi_{eL+}$,
which is the even component of the 5D Dirac spinor $\Psi_{eL}= \Psi_{eL+} 
+ \Psi_{eL-}$. The odd component $\Psi_{eL-}$ does not have a zero mode, 
but at the massive level it pairs up with the massive modes of $\Psi_{eL+}$ 
to form a vectorlike Dirac mass. This embedding of 4D fermions into 5D 
fermions is repeated for each Standard Model fermion. The Standard Model
Yukawa interactions, such as ${\bar \Psi}_{eL} \Psi_{eR} H$, are then 
promoted to 5D interactions in the warped bulk. This gives
\begin{eqnarray}
&& \int d^4 x \int dy~\sqrt{-g}~\lambda_{ij}^{(5)}\left[ 
      \bar\Psi_{iL}(x,y) \Psi_{jR}(x,y) + h.c.\right]
      H(x)\delta(y-\pi R\nonumber)\\
&&      \equiv\int d^4 x~\lambda_{ij}~(\bar\Psi_{iL+}^{(0)}(x) 
      \Psi_{jR+}^{(0)}(x)H(x) + h.c. + \dots)~,
\end{eqnarray}
where $i,j$ are flavour indices, $\lambda_{ij}^{(5)}$ is the (dimensionful) 
5D Yukawa coupling and $\lambda_{ij}$ is the (dimensionless) 
4D Yukawa coupling. Given that the zero mode profile is
\begin{equation}
   \widetilde\psi^{(0)}_{iL+,R+}(y)\propto e^{(\frac{1}{2}-c_{iL,R})k y}~,
\end{equation}
this leads to an exponential hierarchy in the 4D Yukawa coupling~\cite{gp}
\begin{equation}
\label{yukcoup}
   \lambda_{ij} \simeq \lambda_{ij}^{(5)} k
    \sqrt{(c_{iL}-1/2)(c_{iR}-1/2)}~e^{(1-c_{iL}-c_{jR})\pi k R}~,
\end{equation}
for  $c_{iL,R} > 1/2$. Assuming $c_{iL}=c_{jR}$ for simplicity then the 
electron Yukawa coupling $\lambda_e\sim 10^{-6}$ is obtained for 
$c_e\simeq 0.64$. Instead when $c_{iL,R} <1/2$, both fermions are localised
near the IR brane giving 
\begin{equation}
   \lambda_{ij} \simeq \lambda_{ij}^{(5)} k 
       \sqrt{(1/2-c_{iL})(1/2-c_{iR})}~,
\end{equation}
with no exponential suppression. Hence 
the top Yukawa coupling $\lambda_t\sim 1$ is obtained for $c_t \simeq -0.5$. 
The remaining fermion Yukawa couplings, $c_f$ then range from 
$c_t < c_f < c_e$~\cite{gp,hs}. Thus, we see that for bulk mass parameters 
$c$ of ${\cal O}(1)$ the fermion mass hierarchy is explained. The fermion 
mass problem is now reduced to determining the $c$ parameters in the 
5D theory. This requires a UV completion of the 5D warped bulk model 
with fermions, such as string theory.

\begin{figure}[ht]
\centerline{ 
  \epsfxsize 2.5  truein 
   \epsfbox {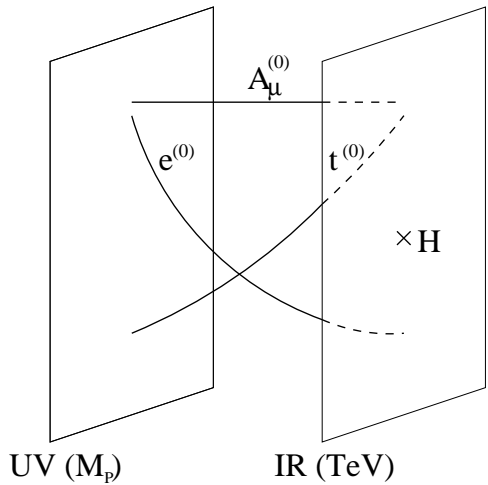}}
\caption{\it The Standard Model in the warped five-dimensional bulk.}
\label{bulkSM}
\end{figure}

Since bulk fermions naturally lead to Yukawa coupling hierarchies,
the gauge bosons must also necessarily be in the bulk. From Table 1
the gauge field zero mode is flat and therefore couples with 
equal strength to both the UV and IR brane. Only the Higgs field is confined 
to the IR brane (or TeV brane). Thus the picture that emerges is a Standard 
Model in the warped bulk as depicted in Figure~\ref{bulkSM}. The fermions
are localised to varying degrees in the bulk with the electron zero mode,
being the lightest fermion, furthest away from the Higgs on the TeV brane
while the top, being the heaviest, is closest to the Higgs. This 
enables one to not only solve the hierarchy problem but also address the 
Yukawa coupling hierarchies.

The warped bulk can also be used to obtain naturally small neutrino masses.
Various scenarios are possible. If the right (left) handed neutrino is 
localised near the UV (IR) brane then a tiny Dirac neutrino mass is 
obtained~\cite{gn}. However this requires that lepton number is 
conserved on the UV brane. Instead in the ``reversed'' scenario one can 
place the right (left) handed neutrino near the IR (UV) brane. In this case 
even though lepton number is violated on the UV brane, the neutrinos will 
still obtain naturally tiny Dirac masses~\cite{tg}.

\subsection{Higher-dimension operators}

Let us consider the following generic four-fermion operators which are
relevant for proton decay and $K-\bar K$ mixing
\begin{equation}
   \int d^4 x \int dy~\sqrt{-g}~\frac{1}{M_5^3} \bar\Psi_i\Psi_j
     \bar\Psi_k\Psi_l \equiv \int d^4 x \frac{1}{M_4^2} 
     \bar\Psi_{i+}^{(0)}\Psi_{j+}^{(0)}\bar\Psi_{k+}^{(0)}\Psi_{l+}^{(0)}~,
\end{equation}
where the effective 4D mass scale $M_4$ for 
$1/2\lappeq c_i \lappeq 1$ is approximately given by\cite{gp}
\begin{equation}
\label{4dscale}
    \frac{1}{M_4^2} \simeq \frac{k}{M_5^3} e^{(4-c_i-c_j-c_k-c_l)\pi kR}~.
\end{equation}
If we want the suppression scale for higher-dimension proton decay operators 
to be $M_4\sim M_P$ then (\ref{4dscale}) requires $c_i\simeq 1$
assuming $k\sim M_5\sim M_P$. Unfortunately for these values of $c_i$ the
corresponding Yukawa couplings would be too small. Nevertheless, 
the values of $c$ needed to explain the Yukawa coupling hierarchies still 
suppresses proton decay by a mass scale larger than the TeV 
scale~\cite{gp,huber}. Thus there is no need to impose 
a discrete symmetry which forbids very large higher-dimension operators.

On the other hand the suppression scale for FCNC processes only needs to be 
$M_4 \gappeq 1000$ TeV. This can easily be achieved for the values of $c$ 
that are needed to explain the Yukawa coupling hierarchies. In fact the 
FCNC constraints can be used to obtain a lower bound on the Kaluza-Klein 
mass scale $m_{KK}$. 
For example Kaluza-Klein gluons can mediate $\Delta S=2$ FCNC processes 
at tree level because the fermions are located at different points. In flat 
space with split fermions this leads to strong constraints 
$m_{KK} \gappeq 25-300$ TeV (with the range depending on whether FCNC 
processes violate CP)~\cite{dpq}. However in warped space for 
$c\gappeq 1/2$, the Kaluza-Klein gauge boson coupling to fermions is 
universal even though fermions are split. This is because in warped space 
the Kaluza-Klein gauge boson wave functions are approximately flat near the 
UV brane. The corresponding bound for warped dimensions 
is $m_{KK} \gappeq 2-30$ TeV. This fact that the Kaluza-Klein gauge bosons
couple universally to spatially separated fermions is akin to a GIM-like 
mechanism in the 5D bulk~\cite{gp}. Thus, warped dimensions 
ameliorate the bounds on the Kaluza-Klein scale. 

\subsection{Higgs as a pseudo Nambu-Goldstone Boson}

So far we have said very little about electroweak symmetry breaking and
the Higgs mass. If the Higgs is confined to the IR (or TeV) brane then the 
tree-level Higgs mass parameter is naturally of order 
$\Lambda_{IR}=\Lambda_{UV} e^{-\pi kR}$. 
Since there are fermions and gauge bosons in the bulk the effects of their
corresponding Kaluza-Klein modes must be sufficiently suppressed. This 
requires $k e^{-\pi kR} \sim {\cal O}(\rm TeV)$ and since in RS1
$\Lambda_{UV} \sim 10 k$ we have  
$\Lambda_{IR}\sim {\cal O}(10~{\rm TeV})$. Consequently a modest amount of 
fine-tuning would be required to obtain a physical Higgs mass of order 100 GeV,
as suggested by electroweak precision data~\cite{higgs}. Clearly, it is 
desirable to invoke a symmetry to keep the Higgs mass naturally lighter than 
the IR cutoff scale, such as the spontaneous breaking of a global 
symmetry\footnote{But this is not the unique possibility. As we will see 
later supersymmetry (instead of a global symmetry) can also be used to 
obtain a light Higgs mass.}.

Motivated by the fact that the dimensional reduction of a 
five-dimensional gauge boson $A_M=(A_\mu,A_5)$ contains a scalar field
$A_5$, one can suppose that the Higgs scalar field is part of 
a higher-dimensional gauge field~\cite{higgsxd}. In a slice of AdS$_5$ the 
$A_5$ terms in the gauge boson kinetic term of the bulk Lagrangian 
(\ref{5daction}) are
\begin{equation}
     -\frac{1}{2}\int d^4 x~dy~e^{-2ky} ((\partial A_5)^2 -2
       \eta^{\mu\nu}\partial_\mu A_5 \partial_5 A_\nu) + \dots~. 
\end{equation}
In particular notice that the higher-dimensional gauge symmetry prevents a 
tree-level mass for $A_5$. However
the zero mode of the $A_5$ scalar field must be localised near the IR brane 
for the hierarchy problem to be solved. The solution
for $A_5$ can be obtained by adding a gauge fixing term that cancels the 
mixed $A_5 A_\nu$ term~\cite{rsch,cnp}. This gives the zero mode 
solution $A_5^{(0)}$ with $y$ dependence $\propto e^{+2 k y}$ and 
substituting back into the action leads to
\begin{equation}
     -\frac{1}{2}\int d^4 x~dy~e^{+2ky} (\partial_\mu A_5^{(0)}(x))^2 
     + \dots~.
\end{equation}
Hence with respect to the flat 5D metric the massless scalar mode 
$A_5^{(0)}$ is indeed localised towards the IR brane and therefore
can play the role of the Higgs boson.

To obtain a realistic model one assumes 
an SO(5)$\times$U(1)$_{B-L}$ gauge symmetry in the 
bulk for the electroweak sector~\cite{cnp}. On the 
IR brane this symmetry is spontaneously broken by boundary conditions to 
SO(4)$\times$U(1)$_{B-L}$. This leads to four Nambu-Goldstone 
bosons that can be identified with the Standard Model Higgs 
doublet. A Higgs mass is then generated because SO(5) gauge 
symmetry is explicitly broken in the fermion sector, in particular by the top 
quark. At one loop this generates an effective potential and electroweak 
symmetry is broken dynamically via top-quark loop corrections~\cite{acp}. 
This effect is finite and arises from the Hosotani mechanism with
nonlocal operators in the bulk~\cite{hosotani}. An unbroken 
SO(3) custodial symmetry guarantees that the Peskin-Takeuichi parameter 
$T=0$. The important point however is 
that radiative corrections to the Higgs mass depend on $k e^{-\pi k R}$ 
and not on $\Lambda_{UV} e^{-\pi kR}$. Together with the accompanying
one-loop factor $\frac{1}{16\pi^2}$ this guarantees a light Higgs mass of 
order $m_{Higgs} \lappeq 140$ GeV.
Furthermore this model can be shown to pass stringent electroweak precision 
tests without a significant amount of fine-tuning~\cite{ac}.

In summary, the hierarchy problem can be solved by placing the 
Standard Model in the warped bulk together with identifying the 
Higgs scalar field as a pseudo Nambu-Goldstone boson. This leads to a very
predictive scenario for the electroweak symmetry breaking sector.
Moreover the setup is radiatively stable and valid up to the Planck scale 
with (as we will see later) grand unification incorporated in an 
interesting way! This makes the 5D warped bulk a compelling alternative 
framework to address the hierarchy problem in a complete scenario compared
to the usual 4D scenarios. But even more compelling is that this 
framework can be given a purely 4D holographic description in terms of a 
strongly coupled gauge theory as will be shown in the next section.

\section {AdS/CFT and holography}
Remarkably 5D warped models in a slice of AdS 
can be given a purely 4D description. This holographic correspondence 
between the 5D theory and the 4D theory originates 
from the AdS/CFT correspondence in string theory. In 1997 Maldacena 
conjectured that~\cite{malda}
\begin{equation}
\begin{tabular}{c}
   {\rm type IIB string theory}\\
     on  AdS$_5\times S^5$     
\end{tabular}
\begin{tabular}{c}
{\footnotesize\sc DUAL}\\[-2mm]
$\quad\Longleftrightarrow\quad$
\end{tabular}
\begin{tabular}{c}
${\cal N}=4$~SU(N) 4D gauge theory 
\end{tabular}
\end{equation}
where ${\cal N}$ is the number of supersymmetry generators and 
\begin{equation}
      \frac{R_{AdS}^4}{l_s^4} = 4\pi g_{YM}^2 N~,
\end{equation}
with $R_{AdS}\equiv 1/k$, $l_s$ is the string length and $g_{YM}$ is the 
SU(N) Yang-Mills gauge coupling. Qualitatively we can see that the isometry 
of the five-dimensional sphere $S^5$ is the rotation group SO(6) $\cong$ 
SU(4), which is the R-symmetry group of the supersymmetric gauge theory. 
Moreover the ${\cal N}=4$ gauge theory is a conformal field theory because the 
isometry group of AdS$_5$ is precisely the conformal group in four 
dimensions. In particular this means that the gauge couplings do not receive 
quantum corrections and therefore do not run with energy. 

In the warped bulk we have only considered gravity. This
represents the effective low energy theory of the full string theory. In 
order to neglect the string corrections, so that the bulk gravity description 
is valid, we require that $R_{AdS}\gg l_s$. This leads to the condition that 
$g_{YM}^2 N \gg 1$, which means that the 4D dual CFT is strongly coupled!
Thus for our purposes the correspondence takes the form of a duality in which 
the weakly coupled  5D gravity description is dual to a strongly coupled 
4D CFT. This remarkable duality means that any geometric configuration of 
fields in the bulk can be given a purely 4D description in terms of a 
strongly coupled gauge theory. Therefore warped models provide a new way 
to study strongly coupled gauge theories.

While there is no rigorous mathematical proof of the AdS/CFT conjecture,
it has passed many nontrivial tests and an AdS/CFT dictionary 
to relate the two dual descriptions can be established.
For every bulk field $\Phi$ there is an associated operator $\cal O$ of the 
CFT. In the AdS$_5$ metric (\ref{adsmetric}) the boundary of AdS space is 
located at $y=-\infty$. The boundary value of the bulk field 
$\Phi(x^\mu,y=-\infty)\equiv \phi_0(x^\mu)$ acts as a source field for 
the CFT operator ${\cal O}$. The AdS/CFT correspondence can then be 
quantified in the following way~\cite{gkp,witten}
\begin{equation}
\label{adsrel}
     \int {\cal D}\phi_{CFT} ~e^{-S_{CFT}[\phi_{CFT}] - \int d^4 x ~\phi_0  
    {\cal O}} = \int_{\phi_0}{\cal D}\phi~ e^{-S_{bulk}[\phi]}
     \equiv e^{i S_{eff}[\phi_0]}~,
\end{equation}
where $S_{CFT}$ is the CFT action with $\phi_{CFT}$ generically denoting the
CFT fields and $S_{bulk}$ is the bulk 5D action. The on-shell gravity action, 
$S_{eff}$ is obtained by integrating out the bulk degrees of freedom for 
suitably chosen IR boundary conditions.
In general $n$-point functions can be calculated via
\begin{equation}
    \langle{ \cal O} \dots {\cal O}\rangle = \frac{\delta^n S_{eff}}
      {\delta\phi_0\dots\delta\phi_0}~.
\end{equation}
In this way we see that the on-shell bulk action is the generating 
functional for connected Greens functions in the CFT. 

So far the 
correspondence has been formulated purely in AdS$_5$ without the presence of 
the UV and IR branes. In particular notice from (\ref{adsrel}) that 
the source field $\phi_0$ is a nondynamical field with no kinetic term. 
However since we are interested in the 4D dual of a slice of AdS$_5$ (and not
the full AdS space) we will need the corresponding dual description in the 
presence of two branes.

Suppose that a UV brane is placed at $y=0$. The $-\infty<y<0$ part
of AdS space is chopped off and the remaining $0< y< \infty$ part is
reflected about $y=0$ with a $Z_2$ symmetry. The presence of the UV brane 
with an associated UV scale $\Lambda_{UV}$ thus corresponds to an explicit 
breaking of the conformal invariance in the CFT at the UV scale 
(but only by nonrenormalisable terms)~\cite{arp,rz,pv}. 
The fact that the CFT now has a finite UV cutoff means that the source field 
$\phi_0$ becomes dynamical. A kinetic term for the source field will
always be induced by the CFT but one can directly add an explicit kinetic 
term for the source field at the UV scale. Thus in the presence of a UV brane 
the AdS/CFT correspondence is modified to the form
\begin{eqnarray}
\label{adsbdyrel}
  \int{\cal D}\phi_0~e^{-S_{UV}[\phi_0]} \int_{\Lambda_{UV} }{\cal D}
    \phi_{CFT} ~e^{-S_{CFT}[\phi_{CFT}] - \int d^4 x ~\phi_0  
     {\cal O}}\nonumber \\
  =\int{\cal D}\phi_0 ~e^{-S_{UV}[\phi_0]}\int_{\phi_0}
    {\cal D}\phi~ e^{-S_{bulk}[\phi]}~,
\end{eqnarray}
where $S_{UV}$ is the UV Lagrangian for the source field $\phi_0$. It is 
understood that now the source field $\phi_0 =\Phi(x,y=0)$. Moving 
away from the UV brane at $y=0$ in the bulk corresponds in the 4D dual to 
running down from the UV scale to lower energy scales. Since the bulk is AdS
the 4D dual gauge theory quickly becomes conformal at energies below the 
UV scale.

The presence of the IR brane at $y=\pi R$ corresponds to a spontaneous 
breaking of the conformal invariance in the CFT at the IR scale 
$\Lambda_{IR}=\Lambda_{UV} e^{-\pi kR}$~\cite{arp,rz}. The conformal 
symmetry 
is nonlinearly realised and particle bound states of the CFT can now appear. 
Formally this can be understood by noting that the (massless) radion field 
in RS1 is localised on the IR brane, since by sending the UV brane to the 
AdS boundary (at $y\rightarrow -\infty$), while keeping the IR 
scale fixed, formally decouples the source field and keeps the radion in 
the spectrum. This means that at the IR scale the CFT must contain a 
massless particle which is interpreted as the Nambu-Goldstone boson of 
spontaneously broken conformal symmetry. This so called dilaton is therefore 
the dual interpretation of the radion. A similar phenomenon also occurs in 
QCD where massless pions are the Nambu-Goldstone bosons of the spontaneously 
broken chiral symmetry at $\Lambda_{QCD}$. Indeed the AdS/CFT correspondence 
suggests that QCD may be the holographic description of a bulk (string) theory.

Thus, the dual interpretation of a slice of AdS not only contains a 4D dual 
CFT with a UV cutoff, but also a dynamical source field $\phi_0$ with UV 
Lagrangian $S_{UV}[\phi_0]$. In particular note that the source field is 
an elementary (point-like) state up to the UV scale, while 
particles in the CFT sector are only effectively point-like below the 
IR scale but are composite above the IR scale. The situation is analogous
to the (elementary) photon of QED interacting with the (composite)
spin-1 mesons of QCD with cutoff scale $\Lambda_{QCD}$.

\subsection{Holography of scalar fields}

As a simple application of the AdS/CFT correspondence in a slice of AdS$_5$
we shall investigate in more detail the dual theory corresponding 
to a bulk scalar field $\Phi$ with boundary mass terms. The qualitative 
features will be very similar for other spin fields. In order to obtain 
the correlation functions of the dual theory we first need to compute the 
on-shell bulk action $S_{eff}$. According to (\ref{genwf}) the bulk scalar 
solution is given by
\begin{equation}
\label{bulkscsoln}
   \Phi(p,z)= \Phi(p) A^{-2}(z)\left[J_\alpha(i q) -\frac{J_{\alpha\pm 1}
    (iq_1)}{Y_{\alpha\pm 1}(i q_1)}Y_\alpha(i q)\right]~,
\end{equation}
where $z=(e^{ky}-1)/k$, $A(z)=(1+kz)^{-1}$, $q=p/(kA(z))$ and $\Phi(p,z)$ is 
the 4D Fourier transform of $\Phi(x,z)$. The $\pm$ refers to the two branches 
associated with $b=b_\pm=2\pm\alpha$. Substituting this solution into the 
bulk scalar action and imposing the IR boundary condition (\ref{scalarbc}) 
leads to the on-shell action
\begin{eqnarray}
\label{effS}
    S_{eff} &=& \frac{1}{2}\int \frac{d^4 p}{(2\pi)^4} \left[
     A^3(z)\Phi(p,z)\left(\Phi'(-p,z)- b~k~A(z)\Phi(-p,z)\right)\right]
     \bigg|_{z=z_0}\nonumber \\
    &=&\frac{k}{2}\int \frac{d^4 p}{(2\pi)^4} F(q_0,q_1) \Phi(p) \Phi(-p)~,
\end{eqnarray}
where 
\begin{eqnarray}
    F(q_0,q_1) &=& \mp~iq_0 \left[ J_{\nu \mp 1}(iq_0) - 
Y_{\nu \mp1} (iq_0)\; \frac{J_{\nu}(iq_1)}{Y_{\nu}(iq_1)}
\right]\nonumber\\
&& \qquad\qquad\qquad\times\left[ J_{\nu}(iq_0) - Y_{\nu} (iq_0)\; 
\frac{J_{\nu}(iq_1)}{Y_{\nu}(iq_1)} \right]~,
\end{eqnarray}
and $\nu\equiv\nu_\pm=\alpha\pm 1$.
\\
\\
\noindent
$\bullet$ {\sc Exercise:} {\it Verify (\ref{effS}) is obtained by
substituting the bulk scalar solution (\ref{bulkscsoln}) 
into the scalar part of the action (\ref{5daction})}.
\\
\\
The dual theory two-point function of the operator $\cal O$ sourced by the 
bulk field $\Phi$ is contained in the self-energy $\Sigma(p)$ obtained by
\begin{eqnarray}
\label{sigp}
      \Sigma(p) &=& \int d^4x\, e^{-ip\cdot x} \frac{\delta^2 S_{eff}}
     {\delta(A^2(z_0) \Phi(x,z_0))\delta(A^2(z_0)\Phi(0,z_0))}~,\nonumber\\
     &=& \frac{k}{g_\phi^2}
     \frac{q_0 (I_\nu(q_0)K_\nu(q_1)-I_\nu(q_1)K_\nu(q_0))}
     {I_{\nu\mp 1}(q_0)K_\nu(q_1)+I_\nu(q_1)K_{\nu\mp 1}(q_0)}~,
\end{eqnarray}
where a coefficient $1/g_\phi^2$ has been factored out in front of the
scalar kinetic term in (\ref{5daction}), so that $g_\phi$ is a 5D expansion 
parameter with dim$[1/g_\phi^2]=1$.
The behaviour of $\Sigma(p)$ can now be studied for various momentum limits
in order to obtain information about the dual 4D theory.
When $A_1\equiv A(z_1)\rightarrow 0$ the effects of the conformal symmetry
breaking (from the IR brane) are completely negligible. The leading 
nonanalytic piece in $\Sigma(p)$ is then interpreted as the pure 
CFT correlator $\langle \cal O \cal O \rangle$ that would be obtained 
in the string AdS/CFT correspondence with 
$A_0\equiv A(z_0)\rightarrow \infty$. However in a slice of AdS the 
poles of $\langle {\cal O}{\cal O}\rangle$ determines the pure CFT
mass spectrum with a nondynamical source field $\phi_0$.
These poles are identical to the poles of $\Sigma(p)$ 
since $\Sigma(p)$ and $\langle \cal O \cal O \rangle$ only differ by 
analytic terms. Hence the poles of the correlator $\Sigma(p)$ correspond to 
the Kaluza-Klein spectrum of the bulk scalar fields 
with Dirichlet boundary conditions on the UV brane.

There are also analytic terms in $\Sigma(p)$. In the string version of the 
AdS/CFT correspondence these terms are subtracted away by adding appropriate 
counterterms. However with a finite UV cutoff (corresponding to the scale
of the UV brane) these terms are now interpreted as kinetic (and higher 
derivative terms) of the source field $\phi_0$, so that the source becomes 
dynamical in the holographic dual theory. The source field can now mix with 
the CFT bound states and therefore the self-energy $\Sigma(p)$ must be resummed
and the modified mass spectrum is obtained by inverting the whole
quadratic term $S_{UV} + S_{eff}$. In the case with no UV boundary action 
$S_{UV}$ this means that the zeroes of (\ref{sigp}) are identical with 
the Kaluza-Klein mass spectrum (\ref{scKK}) corresponding to (modified) 
Neumann conditions for the source field. In both cases (either Dirichlet or
Neumann) these results are consistent with the fact that the 
Kaluza-Klein states are identified with the CFT bound states.
\\
\\
\noindent
$\bullet$ {\sc Exercise:} {\it Check that the zeroes of (\ref{sigp})
agree with the Kaluza-Klein spectrum (\ref{scKK}). }
\\

At first sight it is not apparent that there are an infinite number of 
bound states in the 4D dual theory required to match the infinite number of 
Kaluza-Klein modes in the 5D theory. How is this possible in the 4D gauge 
theory? It has been known since the early 1970's that the two-point 
function in large-$N$ QCD can be written as~\cite{thooft, wittenN}
\begin{equation}
    \langle {\cal O}(p){\cal O}(-p)\rangle = \sum_{n=1}^\infty 
      \frac{F_n^2}{p^2+m_n^2}~,
\end{equation}
where the matrix element for ${\cal O}$ to create the $n$th meson with mass
$m_n$ from the vacuum is $F_n=\langle 0| {\cal O}|n \rangle \propto 
\sqrt{N}/(4\pi)$. In the large $N$ limit the intermediate states are one-meson 
states and the sum must be infinite because we know that the two-point 
function behaves logarithmically for large $p^2$. Since the 4D 
dual theory is a strongly-coupled SU(N) gauge theory that is conformal 
at large scales, 
it will have this same behaviour. This clearly has the same qualitative 
features as a Kaluza-Klein tower and therefore a dual 5D interpretation 
could have been posited in the 1970's!

To obtain the holographic interpretation of the bulk scalar field, recall
that the scalar zero mode can be localised anywhere in the bulk with 
$-\infty <b< \infty$ where $b\equiv b_\pm=2\pm \alpha$ and 
$-\infty < b_- < 2$ and $2 < b_+ <\infty$. Since $b_\pm =1\pm \nu_\pm$ 
we have $-1<\nu_-<\infty$
and $1<\nu_+<\infty$. The $\nu_-$ branch corresponds to $b_- <2$, while
the $\nu_+$ branch corresponds to $b_+> 2$. Hence the $\nu_- (\nu_+)$ branch 
contains zero modes which are localised on the UV (IR) brane.

\subsubsection{$\nu_-$ branch holography}
We begin first with the $\nu_-$ branch. In the limit
$A_0 \rightarrow\infty$ and $A_1\rightarrow 0$ one obtains
\begin{equation}
\label{bmsig}
    \Sigma(p)\simeq -\frac{2k}{g_\phi^2}\left[\frac{1}{\nu}
     \left(\frac{q_0}{2}\right)^2 + \left(\frac{q_0}{2}\right)^{2\nu+2} 
       \frac{\Gamma(-\nu)}{\Gamma(\nu+1)}+\dots\right]~,
\end{equation}
where the expansion is valid for noninteger $\nu$. The expansion for integer
$\nu$ will contain logarithmns. Only the leading analytic term has been 
written in (\ref{bmsig}). The nonanalytic term
is the pure CFT contribution to the correlator $\langle \cal O\cal O\rangle$.
Formally it is obtained by rescaling the fields by an amount $A_0^{\nu+1}$
and taking the limit
\begin{equation}
   \langle {\cal O}{\cal O}\rangle = \lim_{A_0\rightarrow\infty} (\Sigma(p) +
    {\rm counterterms})=\frac{1}{g_\phi^2}\frac{\Gamma(-\nu)}{\Gamma(\nu+1)} 
     \frac{p^{2(\nu+1)}}{(2k)^{2\nu+1}}~.
\end{equation}
Since
\begin{equation}
   \langle {\cal O}(x) {\cal O}(0)\rangle = \int \frac{d^4p}{(2\pi)^4} 
     e^{ipx}\langle {\cal O} {\cal O}\rangle~,
\end{equation}
the scaling dimension of the operator $\cal O$ is
\begin{equation}
    {\rm dim}\,{\cal O} = 3+\nu_- = 4-b_-=2+\sqrt{4+a}~,
\end{equation}
as shown in Table~\ref{tab1}. If $A_0$ is finite then the analytic term in 
(\ref{bmsig}) becomes the kinetic term for the source field $\phi_0$. Placing
the UV brane at $z_0=0$ with $A_0=1$ leads to the dual Lagrangian below the
cutoff scale $\Lambda \sim k$
\begin{equation}
\label{numL}
    {\cal L}_{4D} = -Z_0 (\partial\phi_0)^2 + 
      \frac{\omega}{\Lambda^{\nu_-}}~\phi_0 {\cal O}+{\cal L}_{CFT}~,
\end{equation}
where $Z_0, \omega$ are dimensionless couplings.
This Lagrangian describes a massless dynamical source field $\phi_0$ 
interacting with the CFT via the mixing term $\phi_0{\cal O}$. 
This means that the mass eigenstate in the dual theory will be a mixture 
of the source field and CFT particle states. The coupling of the mixing 
term is irrelevant for $\nu_- >0~(b_-<1)$, marginal if $\nu_-=0~(b_-=1)$ 
and relevant for $\nu_-<0~(b_->1)$. This suggests the following dual 
interpretation of the massless bulk zero mode. When the coupling is irrelevant 
($\nu_- >0$), corresponding to a UV brane localised bulk zero mode,
the mixing can be neglected at low energies, and hence to a very good 
approximation the bulk zero mode is dual to the massless 4D source field 
$\phi_0$. However for relevant $(-1 <\nu_- <0)$ or marginal couplings 
$(\nu_-=0)$ the mixing can no longer be neglected. In this case the bulk 
zero mode is no longer UV-brane localised, and the dual interpretation of 
the bulk zero mode is a part elementary, part composite mixture of the 
source field with massive CFT particle states. 

The first analytic term in (\ref{bmsig}) can be matched to the wavefunction
constant giving $Z_0 = 1/(2\nu g_\phi^2 k)$. However at low energies 
the couplings in ${\cal L}$ will change. The low energy limit $q_1\ll 1$ 
for $\Sigma(p)$ (and noninteger $\nu$) leads to
\begin{equation}
    \Sigma(p)_{IR}\simeq -\frac{2k}{g_\phi^2}\left[(1-A_1^{2\nu_-})
    \left(\frac{q_0}{2}\right)^2\frac{1}{\nu}+\ldots\right]~,
\label{irsig}
\end{equation}
where $A_1=e^{-\pi kR}$. Notice that there is no nonanalytic term because 
the massive CFT modes have decoupled. The analytic term has now also 
received a contribution from integrating out the massive CFT states. 
Note that when $\nu_->0$ the $A_1$ contribution to $Z_0$ is negligible 
and the kinetic term has the correct sign. On the other hand for relevant 
couplings the $A_1$ term now dominates the $Z_0$ term. The kinetic term
still has the correct sign because $\nu_- <0$.
\\
\\
$\bullet$ {\sc Exercise:} {\it Show that for a marginal coupling 
($\nu_-=0$) the coefficient of the kinetic term is logarithmic and has the
correct sign.}
\\
\\
The features of the couplings in (\ref{numL}) at low energies can be neatly 
encoded into a renormalisation group equation. If we define a dimensionless 
running coupling $\xi(\mu) = \omega/\sqrt{Z(\mu)}(\mu/\Lambda)^{\gamma}$, 
which represents the mixing between the CFT and source sector with a 
canonically normalised kinetic term, then it will 
satisfy the renormalisation group equation~\cite{cp}
\begin{equation}
\label{rge}
    \mu\frac{d\xi}{d\mu} = \gamma~\xi + \eta
      \frac{N}{16\pi^2}\xi^3+\dots~,
\end{equation}
where $\eta$ is a constant and we have replaced $1/(g_\phi^2 k) = 
N/(16\pi^2)$. The first term arises from the scaling of the coupling
of the mixing term $\phi_0{\cal O}$ (i.e. $\gamma =\nu_-$), 
and the second term arises from the CFT contribution to the wavefunction 
constant $Z_0$ (i.e. the second term in (\ref{bmsig})). The solution of the 
renormalisation group equation for an initial condition $\xi(M)$ at the scale 
$M\sim\Lambda$ is
\begin{equation}
\label{rgesoln}
     \xi(\mu)= \left(\frac{\mu}{M}\right)^\gamma
     \left\{\frac{1}{\xi^2(M)}+\eta\frac{N}{16\pi^2\gamma}
     \left[1-\left(\frac{\mu}{M}\right)^{2\gamma}\right]\right\}^{-1/2}~.
\end{equation}
When $\gamma<0$, the constant $\eta>0$ and the renormalisation group equation 
(\ref{rge}) has a fixed point at $\xi_\ast \sim 4\pi\sqrt{-\gamma/(\eta N)}$,
which does not depend on the initial value $\xi(M)$. This occurs when 
$-1<\nu_-< 0$ and therefore since  $\xi_\ast$ is nonnegligible the mixing 
between the source and the CFT cannot be neglected. 

In the opposite limit, $\gamma>0$, the solution (\ref{rgesoln}) for 
$M\sim\Lambda$ becomes $\xi(\mu)\sim 4\pi\sqrt{\gamma/N}(\mu/M)^\gamma$, 
where the solution (\ref{rgesoln}) has been matched to the low energy value 
$Z(ke^{-\pi kR})=1/(2\gamma g_\phi^2 k)(1-e^{-2\gamma\pi kR})$ arising from 
(\ref{irsig}) (with $\gamma=\nu_-$). Thus when $\nu_- >0$ the mixing between 
the source and CFT sector quickly becomes irrelevant at low energies. 

\subsubsection{$\nu_+$ branch holography}
Consider the case $\nu=\nu_+ >1$. In the limit 
$A_0 \rightarrow\infty$ and $A_1\rightarrow 0$ we obtain for noninteger $\nu$
\begin{equation}
\label{apUVcor}
 \Sigma(p)\simeq -\frac{2k}{g_\phi^2}\left[ (\nu-1) +  \left(\frac{q_0}{2}
  \right)^2\frac{1}{(\nu-2)}+ \left(\frac{q_0}{2}\right)^{2\nu-2}
  \frac{\Gamma(2-\nu)}{\Gamma(\nu-1)}\right]\nonumber~,
\end{equation}
where only the leading analytic terms have been written. The nonanalytic
term is again the pure CFT contribution to the correlator 
$\langle {\cal O}{\cal O}\rangle$ and gives rise to the scaling
dimension
\begin{equation}
   {\rm dim}\,{\cal O} = 1+\nu_+ = b_+= 2+\sqrt{4+a}~.
\end{equation}
This agrees with the result for the $\nu_-$ branch. At low energies 
$q_1\ll 1$ one obtains
\begin{equation}
\label{IRcorr}
  \Sigma(p)_{IR} \simeq -\frac{2k}{g_\phi^2}\left[ (\nu-1) + 
   \left(\frac{q_0}{2}\right)^2 \frac{1}{(\nu-2)}  - \nu(\nu-1)^2 \; 
   \frac{A_1^{2\nu}}{A_0^{2\nu}}\,\left(\frac{2}{q_0}\right)^2\right]~,
\end{equation}
where the large-$A_0$ limit was taken first. 
We now see that at low energies the nonanalytic term has a pole at $p^2=0$
with the correlator
\begin{equation}
   \langle {\cal O}{\cal O}\rangle = \frac{8k^3}{g_\phi^2}\nu_+(\nu_+-1)^2
    e^{-2\nu_+\pi kR}\frac{1}{p^2}~,
\end{equation}
where $A_0=1$ and $A_1=e^{-\pi kR}$. This pole indicates that the CFT has 
a massless scalar mode at low energies! What about the massless 
source field? As can be seen from (\ref{apUVcor}) and (\ref{IRcorr}) the 
leading analytic piece is a constant term which corresponds to a mass term 
for the source field~\cite{pv}. This leads to the dual Lagrangian below
the cutoff scale $\Lambda\sim k$
\begin{equation}
      {\cal L}_{4D} = -{\widetilde Z}_0(\partial\phi_0)^2 + m_0^2 \phi_0^2+ 
    \frac{\chi}{\Lambda^{\nu_+-2}} \phi_0 {\cal O} +{\cal L}_{CFT}~,
\end{equation}
where ${\widetilde Z}_0,\chi$ are dimensionless parameters and $m_0$ is a mass 
parameter of order the curvature scale $k$. The bare parameters 
${\widetilde Z}_0$ and $m_0$ can be determined from (\ref{apUVcor}).
Thus, the holographic interpretation is perfectly consistent. There is a 
massless bound state in the CFT and the source field $\phi_0$ 
receives a mass of order the curvature scale and decouples. In the bulk the 
zero mode is always localised towards the IR brane. Indeed for $\nu_+>2$ the 
coupling between the source field and the CFT is irrelevant and therefore the 
mixing from the source sector is negligible. Hence to a good approximation 
the mass eigenstate is predominantly the massless CFT bound state. When 
$1\leq \nu_+ \leq 2$ the mixing can no longer be neglected and the mass 
eigenstate is again part elementary and part composite.

\subsection{Dual 4D description of the Standard Model in the Bulk}
The qualitative features of the scalar field holographic picture can now be
used to give the 4D dual description of the Standard Model gauge fields 
and matter in the 5D bulk. In general for every bulk zero mode field there 
is a corresponding massless eigenstate in the dual 4D theory that is a mixture 
of (elementary) source and (composite) CFT fields. If the bulk zero mode is 
localised towards the UV brane, then in the dual theory the massless 
eigenstate is predominantly the source field. For example, as in RS1
the graviton zero mode is localised towards the UV brane so that in the dual 
theory the massless eigenstate is mostly composed of the graviton source 
field. 

On the other hand the dual interpretation of a bulk zero mode localised 
towards the TeV brane is a state that is predominantly a CFT bound state.
In this instance the source field obtains a mass of order the curvature
scale and decouples from the low energy theory. Depending on the degree
of localisation the bound state mixes with the massive source field. Only 
in the limit where the mode is completely localised on the TeV brane is 
the dual eigenmode a pure CFT bound state. Since the Higgs is confined to 
the IR brane, the Higgs field is interpreted as a pure bound state of the CFT 
in the dual theory. In this way we see that the RS1 solution to the hierarchy 
problem is holographically identical to the way 4D composite models solve the 
problem with a low-scale cutoff.
The Higgs mass is quadratically divergent but only sensitive to the strong 
coupling scale $\Lambda_{IR}$ which is hierarchically
smaller than $\Lambda_{UV}$. To obtain a large top Yukawa coupling
the top quark was localised near the IR brane, so in the dual theory the
top quark is also (predominantly) a composite of the CFT. The rest of the 
light fermions are localised to varying degrees towards the UV brane, and
therefore these states will be mostly elementary particle states in the dual
theory. The detailed holographic picture of bulk fermions can be found in
Ref.~\cite{cp}.

If zero modes are not localised in the bulk then the corresponding
4D dual massless eigenstate is partly composed of the elementary source field
and the composite CFT state. In particular for bulk gauge fields whose
zero modes are not localised in the bulk, the 4D dual massless eigenstate
will be part composite and part elementary. 
Finally note that local symmetries in the bulk, such as gauge symmetries 
or general coordinate invariance, also have a 4D dual interpretation. 
The holographic dual of a local symmetry group $G$ in the bulk is a CFT 
in which a subgroup $G$ of the global symmetry group of the CFT is weakly
gauged by the source gauge field~\cite{arp,ad}.

Thus, in summary the Standard Model in the warped 5D bulk is dual to a 
4D hybrid theory with a mixture of elementary and composite states.

\subsubsection{Yukawa couplings}
\begin{figure}[ht]
\centerline{ 
  \epsfxsize 1.5  truein 
   \epsfbox {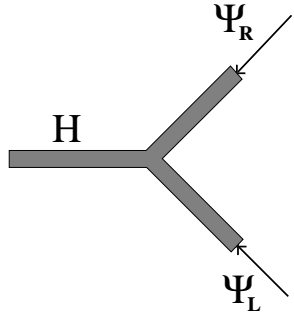}}
\caption{\it The three-point Yukawa coupling vertex in the dual theory
when the fermions are predominantly elementary source fields.}
\label{Yukcoup}
\end{figure}

The Yukawa coupling hierarchies can also be understood from the dual 4D theory.
Consider first an electron (or light fermion) with $c>1/2$. In the dual 4D 
theory the electron is predominantly an elementary field. The dual 4D 
Lagrangian is obtained from analysing $\Sigma(p)$ for fermions, where the CFT
induces a kinetic term for the source field $\psi_L^{(0)}$. 
It is given by~\cite{cp}
\begin{equation}
\label{dualfermL}
     {\cal L}_{4D} = {\cal L}_{CFT} + Z_0 \bar\psi_L^{(0)} i\gamma^\mu
     \partial_\mu\psi_L^{(0)} + \frac{\omega}{\Lambda^{|c+\frac{1}{2}|-1}} 
     (\bar\psi_L^{(0)} {\cal O}_R +h.c.)~,
\end{equation}
where $Z_0, \omega$ are dimensionless couplings and 
dim ${\cal O}_R = 3/2 +|c+1/2|$. The source field $\psi_L^{(0)}$ pertains to 
the left-handed electron $e_L$ and a similar Lagrangian is written for the 
right-handed electron $e_R$. At energy scales $\mu<k$ we have 
a renormalisation group equation like (\ref{rge}) for the mixing parameter
$\xi$ but with $\gamma=|c+1/2|-1$.
Since $c>1/2$ the first term in (\ref{rge}) dominates and the coupling 
$\xi$ decreases in the IR. In particular at the TeV scale ($ke^{-\pi kR}$) 
the solution (\ref{rgesoln}) gives
\begin{equation}
     \xi({\rm TeV})\sim \sqrt{c-\frac{1}{2}} \frac{4\pi}{\sqrt{N}} 
    \left(\frac{k e^{-\pi kR}}{k}\right)^{c-\frac{1}{2}}
    = \sqrt{c-\frac{1}{2}} \frac{4\pi}{\sqrt{N}}e^{-(c-\frac{1}{2})\pi kR}~.
\end{equation}
The actual physical Yukawa coupling $\lambda$ follows from the three-point 
vertex between the physical states. Since both $e_L$ and $e_R$ are
predominantly elementary they can only couple to the composite Higgs
via the mixing term in (\ref{dualfermL}). This is depicted in 
Fig.~\ref{Yukcoup}. In a large-$N$ gauge theory the matrix
element $\langle 0| {\cal O}_{L,R}|\Psi_{L,R}\rangle \sim \sqrt{N}/(4\pi)$,
and the vertex between three composite states 
$\Gamma_3\sim 4\pi/\sqrt{N}$~\cite{wittenN}.
Thus if each of the elementary fields $e_L$ and $e_R$ mixes in the same way 
with the CFT so that $c_{eL}=c_{eR}\equiv c$ then
\begin{equation}
     \lambda \propto  \langle 0|{\cal O}_{L,R}| 
       \Psi_{L,R}\rangle^2~\Gamma_3~\xi^2({\rm TeV})
    = \frac{4\pi}{\sqrt{N}} (c-1/2) e^{-2(c-\frac{1}{2})\pi kR}~.
\end{equation}
This agrees precisely with the bulk calculation (\ref{yukcoup})
where $\lambda^{(5)}_{ij} k \sim 4\pi/\sqrt{N}$.
\\
\\
$\bullet$ {\sc Exercise:} {\it Show that when $|c|<1/2$ the coupling 
of the mixing term in (\ref{dualfermL}) is relevant and the renormalisation 
group equation (\ref{rge}) has a fixed point. Compare this with the
bulk calculation following (\ref{yukcoup}).}
\\
\\
Similarly we can also obtain the Yukawa coupling for the top quark
with $c\lappeq -1/2$ in the dual theory. With this value of $c$ the
top quark is mostly a CFT bound state in the dual theory and we can neglect
the mixing coupling with the CFT. As in the scalar field example this follows 
from the fact that the two point function 
$\langle {\cal O}_R \bar{\cal O}_R\rangle$ now has a massless pole.
The CFT will again generate a mass term for the massless source field, so 
that the only massless state in the dual theory is the CFT bound state. 
The dual Lagrangian is given by~\cite{cp}
\begin{eqnarray}
    {\cal L}_{4D} &=& {\cal L}_{CFT} 
    + Z_0~\bar\psi_L^{(0)} i\gamma^\mu\partial_\mu\psi_L^{(0)} 
    + {\tilde Z}_0~\bar\chi_R i\gamma^\mu\partial_\mu\chi_R\nonumber\\
    &&~+ d~k~({\bar\chi}_R \psi_L^{(0)} + h.c. )
    +~\frac{\omega}{\Lambda^{|c+\frac{1}{2}|-1}}~({\bar\psi}_L^{(0)} 
    {\cal O}_R +h.c.)~,
\end{eqnarray}
where $Z_0, {\tilde Z}_0, d, \omega$ are dimensionless constants.
The fermion $\psi_L^{(0)}$ pertains to $t_L$ and a similar Lagrangian is 
written for $t_R$. Just as in the scalar case this dual Lagrangian is 
inferred from the behaviour of $\Sigma(p)$ for fermions.
The CFT again induces a kinetic term for the source field $\psi_L^{(0)}$ but 
also generates a Dirac mass term of order the curvature scale $k$ with a 
new elementary degree of freedom $\chi_R$. Hence the elementary source field 
decouples from the low energy spectrum and the mixing term is no longer
relevant for the Yukawa coupling. Instead the physical Yukawa coupling
will arise from a vertex amongst three composite states so that
$\lambda_t \sim \Gamma_3 \sim 4\pi/\sqrt{N} \sim \lambda^{(5)} k$, and 
consequently there is no exponential suppression in the Yukawa coupling. 
This is again consistent with the bulk calculation.

\subsubsection{Minimal Composite Higgs model}
Similarly the Higgs as a pseudo Nambu-Goldstone boson scenario has a 
4D dual interpretation~\cite{acp}. Since $A_5$ is localised near the 
TeV-brane the Higgs boson is a composite state in the dual theory. 
The bulk SO(5) gauge symmetry is interpreted as an SO(5) {\it global} 
symmetry of the CFT. This global symmetry is then spontaneously broken down 
to SO(4) at the IR scale by the (unknown) strong dynamics of the CFT. This 
leads to a Nambu-Goldstone boson transforming as a ${\bf 4}$ of SO(4), which 
is a real bidoublet of SU(2)$_L\times$SU(2)$_R$. 

To break electroweak symmetry, an effective Higgs potential is generated 
at one loop by explicitly breaking the SO(5) symmetry in the elementary 
(fermion) sector and transmitting it to the CFT. The top quark plays the 
major role in breaking this symmetry. However the top quark must also
be localised near the IR brane to obtain a large overlap with the Higgs 
field and therefore a large top mass. Moreover to prevent large deviations
from the Standard Model prediction for $Z\rightarrow {\bar b}_L b_L$ the 
left-handed top quark must be localised towards the UV brane 
($c_{tL}\lappeq 1/2$)\cite{adms,acp}. Thus to obtain a large top mass 
the right-handed top quark must be localised near the IR brane.
This specific localisation is compatible with electroweak
symmetry breaking where the Higgs field develops a vacuum expectation value
and breaks SO(4) down to the custodial group SO(3). Hence in
the dual theory the physical right-handed (left-handed) 
top quark is mostly composite (elementary) 
and the custodial symmetry prevents large contributions to the $T$ parameter.

In summary this 4D composite Higgs model is very predictive, with minimal
particle content and is consistent with electroweak precision 
tests~\cite{ac}. This hybrid 4D theory with elementary and composite 
states successfully addresses the hierarchy problem, fermion masses and 
flavour problems in a complete framework.

\section{Supersymmetric Models in Warped Space}
Supersymmetry elegantly solves the hierarchy problem because quadratic
divergences to the Higgs mass are automatically cancelled thereby
stabilising the electroweak sector. However this success must be tempered
with the fact that supersymmetry has to be broken in nature. 
In order to avoid reintroducing a fine-tuning in the Higgs mass, 
the soft mass scale cannot be much larger than the TeV scale. Hence
one needs an explanation for why the supersymmetry breaking scale is low.
Since in warped space hierarchies are easily generated, the warp factor 
can be used to explain the scale of supersymmetry breaking, instead of 
the scale of electroweak breaking. This is one motivation for studying 
supersymmetric models in warped space. Thus, new possibilities 
open up for supersymmetric model building, and in particular for the 
supersymmetry-breaking sector. Moreover by the AdS/CFT correspondence these 
new scenarios have an interesting blend of supersymmetry and compositeness 
that lead to phenomenological consequences at the LHC.

A second motivation arises from the fact that electroweak precision data
favours a light (compared to the TeV scale ) Higgs boson mass~\cite{higgs}. 
As noted earlier the Higgs boson mass in a generic warped model without any 
symmetry is near the IR cutoff (or from the 4D dual perspective the Higgs 
mass is near the compositeness scale). Introducing supersymmetry 
provides a simple reason for why the Higgs boson mass is light and below the 
IR cutoff of the theory.

\subsection{Supersymmetry in a Slice of AdS}
It is straightforward to incorporate supersymmetry in a slice of
AdS~\cite{gp,bagger}. The amount of supersymmetry allowed in five dimensions 
is determined by the dimension of the spinor representations.
In five dimensions only Dirac fermions are allowed by the Lorentz algebra,
so that there are eight supercharges which corresponds from the 4D
point of view to an ${\cal N}=2$ supersymmetry. This means that all bulk 
fields are in ${\cal N}=2$ representations. At the massless level only half 
of the supercharges remain and the orbifold breaks the bulk supersymmetry to 
an ${\cal N}=1$ supersymmetry. 

Consider an ${\cal N}=1$ (massless) chiral multiplet 
$(\phi^{(0)},\psi^{(0)})$ in the bulk. We have seen that the zero mode 
bulk profiles of $\phi^{(0)}$ 
and $\psi^{(0)}$ are parametrised by their bulk mass parameters $a$ and $c$, 
respectively. Since supersymmetry treats the scalar and fermion components 
equally, the bulk profiles of the component fields must be the same. 
It is clear that in general this is 
not the case except when $1\pm\sqrt{4+a} = 1/2 -c$, as follows from the 
exponent of the zero mode profiles in Table~\ref{tab1}. This leads to the
condition that 
\begin{equation}
\label{acond}
   a=c^2+c-15/4~, 
\end{equation}
and the one remaining mass parameter $c$
determines the profile of the chiral multiplet to be
\begin{equation}
      \left(\begin{tabular}{c} $\phi^{(0)}$\\ $\psi^{(0)}$
            \end{tabular}\right) \propto e^{(\frac{1}{2}-c)k y}~.
\end{equation}
Thus for $c > 1/2$ ($c< 1/2$) the chiral supermultiplet is localised 
towards the UV (IR) brane. It can be shown that the scalar boundary mass,
that was tuned to be $b=2\pm\alpha$, follows from the invariance under a 
supersymmetry transformation~\cite{gp} when (\ref{acond}) is satisfied.

Similarly a gauge boson with bulk profile $A_\mu^{(0)}(y) \propto 1$ 
and a gaugino with bulk profile 
$\lambda^{(0)}(y)\propto e^{(\frac{1}{2}-c_\lambda)ky}$ can be combined into
an ${\cal N}=1$ vector multiplet only for $c_\lambda = 1/2$. Of course
this means that the gaugino zero-mode profile is flat like the gauge boson.
At the massive level the on-shell field content of an ${\cal N}=2$ vector
multiplet is $(A_M, \lambda_i,\Sigma)$ where $\lambda_i$ is a 
symplectic-Majorana spinor (with $i=1,2$) and $\Sigma$ is a real scalar in 
the adjoint representation of the gauge group. Invariance under
supersymmetry transformations requires that $\Sigma$ have bulk and boundary
mass terms with $a=-4$ and $b=2$, respectively. So, if $\Sigma$ is even
under the orbifold symmetry then these values will ensure a scalar 
zero mode.

Finally a graviton with bulk profile $h_{\mu\nu}^{(0)}(y)\propto e^{-k y}$ and 
a gravitino with bulk profile $\psi_\mu^{(0)}(y)\propto 
e^{(\frac{1}{2}-c_\psi)ky}$ can be combined into
an ${\cal N}=1$ gravity multiplet only for $c_\psi = 3/2$. In this case 
the gravitino zero-mode profile is localised on the UV brane.

\subsection{The Warped MSSM}

In the warped MSSM the warp factor is used to naturally generate TeV scale 
soft masses~\cite{gp1}. The UV (IR) scale is identified with the 
Planck (TeV) scale. The IR brane is associated with the scale of 
supersymmetry breaking, while the bulk and UV brane are supersymmetric. 
At the massless level the particle content is identical to the 
MSSM. The matter and Higgs superfields are assumed to be confined on the UV 
brane. This naturally ensures that all higher-dimension operators associated 
with proton decay and FCNC processes are sufficiently suppressed. In the bulk
there is only gravity and the Standard Model gauge fields. These are 
contained in an ${\cal N}=1$ gravity multiplet and vector multiplet,
respectively.

Supersymmetry is broken by choosing different IR brane boundary conditions 
between the bosonic and fermionic components of the bulk superfields. On the 
boundaries the condition (\ref{fermpar}) defines a chirality since 
$(1\mp\gamma_5)\Psi(y^\ast) = 0$ where $y^\ast=0$ or $\pi R$. If opposite
chiralities are chosen on the two boundaries then this leads to antiperiodic 
conditions for the fermions, namely
\begin{equation}
    \Psi(y + 2\pi R)= -\Psi(y)~.
\end{equation}
If the gauginos in the bulk are assumed to have opposite chiralities
on the two boundaries then supersymmetry will be broken because the 
gauge bosons obey periodic boundary conditions. The gaugino zero mode
is no longer massless and receives a mass
\begin{equation}
\label{gaugem}
    m_\lambda \simeq \sqrt{\frac{2}{\pi kR}}~k~e^{-\pi kR}~.
\end{equation}
Since the theory has a U(1)$_R$ symmetry this is actually a Dirac mass 
where the gaugino zero mode pairs up with a Kaluza-Klein mode~\cite{gp1}.
The Kaluza-Klein mass spectrum of the gauginos also shifts relative to 
that of the gauge bosons by an amount $-\frac{1}{4}\pi k e^{-\pi kR}$.
Similarly for the gravity multiplet the gravitino is assumed to
have antiperiodic boundary conditions, while the graviton has periodic
boundary conditions. The gravitino zero mode then receives a mass
\begin{equation}
\label{gravm}
    m_{3/2} \simeq \sqrt{8}~k~e^{-2\pi kR}~,
\end{equation}
while the Kaluza-Klein modes are again shifted by an amount similar to
that of the vector multiplet. 
\\
\\
$\bullet$ {\sc Exercise:} {\it Using the expressions (\ref{genwf})
and (\ref{balpha}), check that imposing
opposite chiralities on the two boundaries for the gaugino and gravitino
leads to the zero mode masses (\ref{gaugem}) and (\ref{gravm}).}
\\
\\
If $k e^{-\pi kR}=$ TeV then the gaugino mass (\ref{gaugem}) is
$m_\lambda \simeq 0.24$ TeV while the gravitino mass (\ref{gravm}) is
$m_{3/2}\simeq 3\times 10^{-3}$ eV. Even though both the gaugino and 
gravitino are bulk fields the difference in their supersymmetry breaking 
masses follows from their coupling to the IR brane, which is where
supersymmetry is broken. The gaugino is not localised in the bulk and 
couples to the IR brane with an ${\cal O}(1)$ coupling. Hence it receives 
a TeV scale mass. On the other hand the gravitino is localised on the UV 
brane and its coupling to the TeV brane is exponentially suppressed. 
This explains why the gravitino mass is much smaller than the gaugino mass.

\begin{figure}[ht]
\centerline{ 
  \epsfxsize 2.5  truein 
   \epsfbox {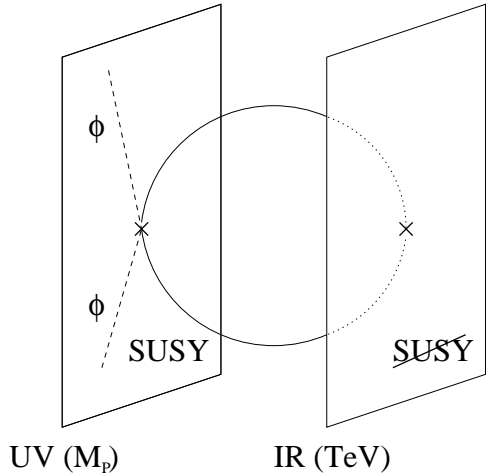}}
\caption{\it The transmission of supersymmetry breaking in the warped MSSM 
to UV-brane localised matter fields via bulk gauge interactions which couple 
directly to the IR brane.}
\label{warpMSSM}
\end{figure}

The scalars on the UV brane will obtain a supersymmetry breaking
mass at one loop via gauge interactions with the bulk vector multiplets. 
The gravity interactions with the gravity multiplet are negligible. 
A one loop calculation leads to the soft mass spectrum
\begin{equation}
\label{softm}
     {\widetilde m}_j^2 \propto \frac{\alpha_i}{4\pi} ({\rm TeV})^2~,
\end{equation}
where $\alpha_i= g_i^2/(4\pi)$ are individual gauge contributions 
corresponding to the particular gauge quantum numbers of the particle state.
The exact expressions are given in Ref.~\cite{gp1}.
Unlike loop corrections to the usual 4D supersymmetric soft masses, 
the masses in (\ref{softm}) are finite. Normally UV divergences in a 
two-point function arise when the two spacetime points coincide. But 
the spacetime points in the 5D loop diagram can never coincide, because 
the two branes are assumed to be a fixed distance apart, and therefore the 
5D one-loop calculation leads to a finite result (see Figure~\ref{warpMSSM}).
This is similar to 
the cancellation of divergences in the Casimir effect~\cite{casimir}. 
Since the contribution to the scalar masses is due to gauge interactions 
the scalar masses are naturally flavour diagonal. This means that the 
right-handed slepton is the lightest scalar particle since it has the smallest
gauge coupling dependence. The lightest supersymmetric particle
will be the superlight gravitino.

\subsubsection{The dual 4D interpretation}
We can use the AdS/CFT dictionary to obtain the dual 4D interpretation 
of the warped MSSM. Clearly the matter and Higgs fields confined
to the UV brane are elementary fields external to the CFT. This is also
true for the zero modes of the gravity multiplet since it is localised 
towards the UV brane. However, the bulk gauge field zero modes 
are partly composite since they are not localised. The Kaluza-Klein states,
which are bound states of the CFT and localised near the IR brane, 
do not respect supersymmetry.
Therefore at the TeV scale not only is conformal symmetry broken by the CFT
but also supersymmetry. This requires some (unknown) nontrivial IR dynamics
of the CFT, but the point is that supersymmetry is dynamically broken.
Since the CFT is charged under the Standard Model gauge group, the gauginos 
(and gravitinos) will receive a tree-level supersymmetry breaking mass,
while the squarks and sleptons will receive their soft mass at one loop.
In some sense this model is very similar to 4D gauge-mediated models
except that there is no messenger sector since the CFT, responsible for
supersymmetry breaking, is charged under the Standard Model gauge group.

In particular the bulk gaugino mass formula (\ref{gaugem}) can be understood 
in the dual theory. Since the gaugino mass is of the Dirac type the gaugino 
(source) field must marry a fermion bound state to become massive. This 
occurs from the mixing term ${\cal L} = \omega \lambda {\cal O}_\psi$.
Since $c_\lambda=1/2$, we have from Table~\ref{tab1} that dim$\,{\cal O}_\psi 
=5/2$ and therefore $\omega$ is dimensionless. This means that the mixing
term coupling runs logarithmically so that at low energies the solution 
of (\ref{rge}) is
\begin{equation}
     \xi^2(\mu)\sim \frac{16\pi^2}{N \log\frac{k}{\mu}}~.
\end{equation}
Thus at $\mu=k e^{-\pi kR}$ we obtain the correct factor in (\ref{gaugem})
since the Dirac mass $m_\lambda \propto \xi\,
\langle 0| {\cal O}_\psi|\Psi\rangle$, where in the large-$N$ limit the 
matrix element for ${\cal O}_\psi$ to create a bound state fermion 
is $\langle 0| {\cal O}_\psi|\Psi\rangle \sim \sqrt{N}/(4\pi)$~\cite{wittenN}.

Thus, in summary we have the dual picture
\begin{equation}
\begin{tabular}{c}
    5D warped\\ MSSM
\end{tabular}
\begin{tabular}{c}
{\footnotesize\sc DUAL}\\[-2mm]
$\quad\Longleftrightarrow\quad$
\end{tabular}
\begin{tabular}{l}
4D MSSM $\oplus$ gravity \\
$\oplus$ strongly coupled 4D CFT
\end{tabular}
\end{equation}

\noindent
The warped MSSM is a very economical model of dynamical
supersymmetry breaking in which the soft mass spectrum is calculable and 
finite, and unlike the usual 4D gauge-mediated models does not require 
a messenger sector. The soft mass TeV scale is naturally explained and the
scalar masses are flavour diagonal. In addition, as we will show later,
gauge coupling unification occurs with logarithmic running~\cite{pomarol} 
arising primarily from the elementary (supersymmetric) sector as in the usual
4D MSSM\cite{gns}.

\subsection{The Partly Supersymmetric Standard Model}

Besides solving the hierarchy problem the supersymmetric standard 
model has two added bonuses. First, it successfully 
predicts gauge coupling unification and second, it 
provides a suitable dark matter candidate. Generically, however, there 
are FCNC and CP violation problems arising from the soft mass Lagrangian, 
as well as the gravitino and moduli problems in cosmology~\cite{pb}. 
These problems stem from the fact that the soft masses are of order 
the TeV scale, as required for a natural solution to the hierarchy problem.
Of course clever mechanisms exist that avoid these problems but 
perhaps the simplest solution would be to have all scalar masses at the 
Planck scale while still naturally solving the hierarchy problem. In the 
partly supersymmetric standard model~\cite{gp2} this is precisely what 
happens while still preserving the successes of the MSSM.

In 5D warped space the setup of the model is as follows. 
Supersymmetry is assumed to be broken on the UV brane while it is 
preserved in the bulk and the IR brane. The vector, matter, and
gravity superfields are in the bulk while the Higgs superfield is 
confined to the IR brane.
On the UV brane the supersymmetry breaking
can be parametrised by a spurion field $\eta=\theta^2 F$, where
$F\sim M_P^2$. In the gauge sector
we can add the following UV brane term
\begin{equation}
      \int d^2\theta~\frac{\eta}{M_P^2}\frac{1}{g_5^2}W^\alpha W_\alpha
     \delta(y) + h.c.
\end{equation} 
This term leads to a gaugino mass for the zero mode $m_\lambda\sim M_P$, 
so that the gaugino decouples from the low energy spectrum. The gravitino
also receives a Planck scale mass via a UV brane coupling 
and decouples from the low energy spectrum~\cite{luty}. Similarly
a supersymmetry breaking mass term for the squarks and sleptons can be added
to the UV brane
\begin{equation}
     \int d^4\theta~\frac{\eta^\dagger\eta}{M_P^4}~k~S^\dagger S~\delta(y)~,
\end{equation}
where $S$ denotes a squark or slepton superfield. This leads to a soft 
scalar mass ${\widetilde m}\sim M_P$, so that the squark and slepton zero 
modes also decouple from the low energy spectrum. 

\begin{figure}[ht]
\centerline{ 
  \epsfxsize 2.5  truein 
   \epsfbox {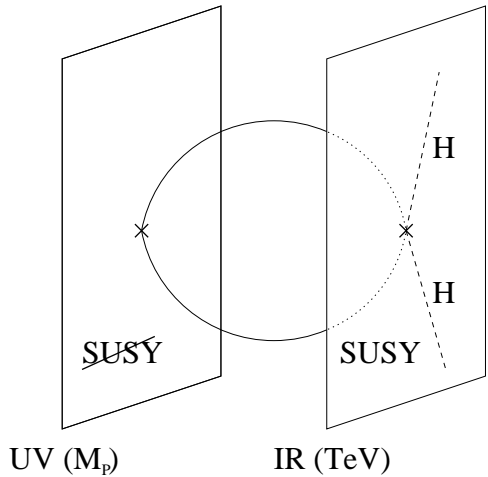}}
\caption{\it The transmission of supersymmetry breaking in the partly
supersymmetric standard model to the supersymmetric Higgs sector via bulk 
gauge interactions which couple directly to the UV brane.}
\label{partSM}
\end{figure}

The Higgs sector is different because the Higgs lives on the IR
brane and there is no direct coupling to the UV brane. Hence, at tree-level
the Higgs mass is zero, but a (finite) soft Higgs mass will be induced at 
one loop via the gauge interactions in the bulk of order
\begin{equation}
     m_H^2 \sim \frac{\alpha}{4\pi} (k e^{-\pi k R})^2 \ll M_P^2~.
\end{equation}
As noted earlier the finiteness is due to the fact that the two 5D spacetime 
points on the UV and IR branes can never coincide (see Figure~\ref{partSM}).
Thus, we see that because of the warp factor 
the induced Higgs soft mass is much smaller than the scale of supersymmetry 
breaking at the Planck scale. So while at the massless 
level the gauginos, squarks and sleptons have received Planck scale masses, 
the Higgs sector remains (approximately) supersymmetric. In summary at the 
massless level the particle spectrum consists of the Standard Model gauge 
fields and matter (quarks and leptons) plus a Higgs scalar and Higgsino. 
This is why the model is referred to as {\it partly} supersymmetric. 

At the massive level the Kaluza-Klein modes are also approximately
supersymmetric. This is because they are localised towards the IR brane 
and have a small coupling to the UV brane. So the Planck scale 
supersymmetry breaking translates into an order TeV scale splitting
between the fermionic and bosonic components of the Kaluza-Klein 
superfields. 

Given that there are no gauginos, squarks or sleptons in the low energy 
spectrum it may seem puzzling how the quadratic divergences cancel in this
model. Normally in the supersymmetric standard model the quadratic 
divergences in the Higgs mass are cancelled by a superpartner contribution 
of the opposite sign. However in the partly supersymmetric standard model
there are no superpartners at the massless level. Instead what happens is
that the difference between the Kaluza-Klein fermions and bosons sums
up to cancel the zero mode quadratic divergence. 
Thus the Kaluza-Klein tower is responsible for keeping the Higgs mass 
natural even though supersymmetry is broken at the Planck scale.

\subsubsection{Higgs sector possibilities}
The motivation for making the Higgs sector supersymmetric is that
the Higgs mass is induced at loop level and therefore the Higgs mass is
naturally suppressed below the IR cutoff. In addition the supersymmetric
partner of the Higgs, the Higgsino, provides a suitable dark matter 
candidate~\cite{gp2,mm}. However since the Higgsino is a fermion,
gauge anomalies could be generated and these must be cancelled.
This leads us to consider the following three possibilities:

{\bf (i) Two Higgs doublets:}
As in the MSSM we can introduce two Higgs doublet superfields $H_1$ and 
$H_2$, so that the gauge anomaly from the two Higgsinos cancel amongst 
themselves. In this scenario we can add the following superpotential on 
the IR brane
\begin{equation}
\label{twoHW}
     \int d^2\theta \left( y_d H_1 Q d + y_u H_2 Q u 
         + y_e H_1 L e +\mu H_1 H_2 \right)~.
\end{equation}
Thus the quarks and leptons receive their masses in the usual way.
In addition the $\mu$ term in (\ref{twoHW}) is naturally of order the 
TeV scale so that there is no $\mu$ problem. The IR brane is approximately
supersymmetric and the supersymmetric mass $\mu$ has a  natural
TeV value. This is unlike the MSSM where the natural scale of $\mu$
is $M_P$ and consequently a problem for phenomenology.

{\bf (ii) One Higgs doublet:}
At first this possibility seems to be ruled since one massless Higgsino
gives rise to a gauge anomaly. However starting with a bulk Higgs
${\cal N}=2$ hypermultiplet $H=(H_1,H_2)$ with bulk mass parameter $c_H=1/2$ 
that consists of two ${\cal N}=1$ chiral
multiplets $H_{1,2}$ we can generate a Higgsino Dirac mass and only
one Higgs scalar doublet in the low energy spectrum. The trick is to use
mixed boundary conditions where $H_1$ has Neumann (Dirichlet) boundary
conditions on the UV (IR) brane and vice versa for $H_2$. This leads
to a $\mu$ term
\begin{equation}
    \mu\simeq \sqrt{\frac{2}{\pi kR}}~k~e^{-\pi kR}~,
\end{equation}
which is similar to the gaugino mass term (\ref{gaugem}) obtained in the 
warped MSSM.
In this case the $\mu$-term is naturally suppressed below TeV scale by
the factor $1/\sqrt{\pi kR}$. Only one Higgs scalar remains in the low 
energy spectrum because the twisted boundary conditions localise
one Higgs scalar doublet towards the UV brane where it obtains a Planck 
scale mass, and the other Higgs scalar is localised towards the IR brane
where it obtains a mass squared $\mu^2$. 

{\bf (iii) Zero Higgs doublet--Higgs as a Slepton:}
No anomalies will occur if the Higgs is considered to be the 
superpartner of the tau (or other lepton). This idea is not
new and dates back to the early days of supersymmetry~\cite{fayet}.
The major obstacle in implementing this possibility in the MSSM is that
the gauginos induce an effective operator $\frac{g^2}{m_\lambda}\nu\nu h h$
that leads to neutrino masses of order 10 GeV which are experimentally
ruled out. However in the partly supersymmetric model $m_\lambda \sim M_P$ 
and neutrinos masses are typically of order $10^{-5}$ eV. This
is phenomenologically acceptable and at least makes this a viable possibility.
However the stumbling block is to generate a realistic 
spectrum of fermion masses without introducing abnormally large
coefficients~\cite{gp2}.

\subsubsection{Electroweak symmetry breaking}
In this model electroweak symmetry breaking can be studied and calculated
using the 5D bulk propagators. Consider, for simplicity, a one Higgs doublet
version of the model. The scalar potential is 
\begin{equation}
\label{higgspot}
    V(h) =\mu^2 |h|^2 + \frac{1}{8} (g^2+g'^2) |h|^4 + V_{gauge}(h) 
    + V_{top}(h)~,
\end{equation}
where $V_{gauge}(h)$ and $V_{top}(h))$ are one-loop contributions to the
effective potential arising from gauge boson and top quark loops, 
respectively. The first two terms in (\ref{higgspot}), which 
arise at tree-level, are monotonically increasing giving rise to a
minimum at $\langle |h| \rangle = 0$ and therefore do not break 
electroweak symmetry. This is why we need to calculate the one-loop 
contributions. The one-loop gauge contribution is given by
\begin{equation}
   V_{gauge}(h) = 6\int_0^\infty \frac{dp}{8\pi^2}~p^3~\log
    \left[\frac{1+ g^2 |h|^2 G_B(p)}{1+ g^2 |h|^2 G_F(p)}\right]~,
\end{equation}
where $G_{B,F}(p)$ are the boson (fermion) gauge propagators in the bulk 
whose expressions can be found in Ref.~\cite{gp2}. The contribution to
the effective potential from $V_{gauge}$ is again monotonically increasing.
However there is also a sizeable contribution from top quark loops (due to
the large top Yukawa coupling) given by
\begin{equation}
   V_{top}(h) = 6\int_0^\infty \frac{dp}{8\pi^2}~p^3~\log
    \left[\frac{1+ p^2 y_t^2 |h|^2 G_B^2(p)}{1+ p^2 y_t^2 |h|^2 G_F(p)}
    \right]~.
\end{equation}
This contribution generates a potential that monotonically decreases with
$|h|$, destabilising the vacuum and thus triggering electroweak symmetry 
breaking. In order for this to occur the top quark needs to be
localised near the IR brane with a bulk mass parameter $c_t \simeq -0.5$.
Since the top quark ${\cal N}=1$ chiral multiplet is localised near the 
IR brane, the top squark will only receive a TeV scale soft mass and
consequently will remain in the low energy spectrum. In fact this radiative
breaking of electroweak symmetry due to a large top Yukawa coupling 
is similar to that occuring in the usual MSSM. As in the MSSM the value of
the Higgs mass is very model dependent but if no large tuning of parameters 
is imposed one obtains a light Higgs boson with mass 
$m_{Higgs}\lappeq 120$ GeV.

\subsubsection{Dual 4D interpretation}
The dual 4D interpretation of the partly supersymmetric model follows
from applying the rules of the AdS/CFT dictionary. 
Supersymmetry is broken at the Planck scale in the dual 4D theory
and is approximately supersymmetric at the IR scale. 
Thus from a 4D point of view supersymmetry is really 
just an accidental symmetry at low energies. At the massless level the 
Higgs is confined on the IR brane and the top quark is localised towards 
the TeV brane so both of these states are CFT composites and supersymmetric
at tree level. The compositeness of the Higgs and stop explains why these 
states are not sensitive to the UV breaking of supersymmetry. These states 
are ``fat'' with a size of order TeV$^{-1}$, and are transparent to
high momenta or short wavelength probes that transmit the breaking of 
supersymmetry. At one loop TeV-scale supersymmetry 
breaking effects arise from the small mixing with the elementary source 
fields, which directly feel the Planck scale supersymmetry breaking. The bulk 
gauge fields are partly composite and the light fermions which are localised 
to varying degrees near the UV brane are predominantly elementary fields. 
Since the light fermion superpartners are predominantly source fields they 
obtain Planck scale soft masses. 

Thus the dual picture can be summarised as follows
\begin{equation}
\begin{tabular}{c}
    5D partly\\ supersymmetric SM
\end{tabular}
\begin{tabular}{c}
{\footnotesize\sc DUAL}\\[-2mm]
$\quad\Longleftrightarrow\quad$
\end{tabular}
\begin{tabular}{l}
4D SM $\oplus$ Higgsino $\oplus$ stop\\
$\oplus$ gravity \\
$\oplus$ strongly coupled 4D CFT
\end{tabular}
\end{equation}

\noindent
The partly supersymmetric standard model is a natural model
of high-scale supersymmetry breaking. Supersymmetry is realised in the
most economical way. Only the Higgs sector and top quark are supersymmetric
and composite, while all other squarks and sleptons have Planck scale masses. 
The Higgsino is the dark matter candidate and even gauge 
coupling unification is achieved as we show in the next section.

\section{Grand Unification}
The unification of the gauge couplings strongly suggests that there is 
an underlying simplifying structure based on grand unified theories 
with gauge groups such as SU(5) and SO(10). We would like to preserve this 
structure in warped extra dimensions as well. At the quantum level
gauge couplings are sensitive to particle states that can be 
excited in the vacuum and this leads to an energy dependence or running.
At one loop the general expression for this energy dependence is 
\begin{equation}
\label{gceq}
     \frac{1}{g_a^2(p)} = \frac{1}{g_U^2} + \frac{b_a}{8\pi^2}
     \log\frac{M_U}{p}~,
\end{equation}
where $g_U$ is the unified gauge coupling, $M_U$ is the unification scale 
and $a=1,2,3$ represents the gauge couplings of U(1)$_Y$, SU(2)$_L$ 
and SU(3), respectively. The coefficients $b_a$ depend on the charged 
particle states that can be excited in the quantum vacuum. At low energies 
the three gauge couplings are different and if these couplings are to unify
at a high energy scale then $b_a$ must be different. From the measurement
of the three gauge couplings at the scale $M_Z$ we can use the three
equations (\ref{gceq}) to obtain the experimental prediction for the 
ratio
\begin{equation}
\label{Bexpt}
    B=\frac{b_3-b_2}{b_2-b_1} = 0.717 \pm 0.008~,
\end{equation}
where the error is due to the experimental values of the gauge 
couplings~\cite{peskin}.
This ratio can also be calculated theoretically for the particle content
of any model. Note that any constant or universal contribution to $b_a$
does not affect the ratio $B$, since it is a ratio of differences.

In the Standard Model the quarks and leptons contribute universally to the 
running because they form complete SU(5) multiplets. Therefore they do not
affect the relative running of the couplings and hence the value of $B$. 
Only the Higgs and Standard Model gauge bosons, which do not form complete
SU(5) multiplets at low energies, affect the relative running. This leads 
to a Standard Model prediction $B=0.528$, which does not agree with the 
experimental value (\ref{Bexpt}), even allowing for a $10\%$ theoretical 
uncertainty due to threshold corrections and higher-loop effects.
 
On the other hand the MSSM doubles the particle spectrum with the addition 
of gauginos, Higgsinos, squarks and sleptons. Again the squarks and sleptons 
form complete SU(5) multiplets and do not contribute to the relative running.
However the extra contributions from the gauginos and Higgsinos combined
with the gauge bosons and Higgs scalar field leads to the MSSM prediction
$B=0.714$. This is remarkably close to the experimental value (\ref{Bexpt}),
even if one accounts for theoretical uncertainties, and is one reason why 
supersymmetry is a leading solution to the hierarchy problem. The question 
we would like to now address is how does gauge coupling unification work for 
models in warped space?

\subsection{Logarithmic Running in 5D Warped Space}
Given that models in 5D warped space can be given a 4D dual description the
gauge couplings should run logarithmically. But how does this happen
given that the 5D model has Kaluza-Klein modes? To answer this
question let us consider the one-loop corrections to the U(1) gauge coupling 
of the zero mode generated by an even bulk 5D scalar $\phi$ with 
charge $+1$~\cite{pomarol}. The Kaluza-Klein spectrum of the scalar is 
given by the expressions (\ref{KKspect}) with $\alpha_\phi=2$.
To regulate the model we introduce a Pauli-Villars (PV) regulator scalar
field $\Phi$ with bulk mass $\Lambda \lappeq k$ and no boundary mass terms. 
This means that the PV field has no massless mode and the zero mode obtains a
mass $m_0 \simeq \Lambda/\sqrt{2}$. On the other hand the Kaluza-Klein 
spectrum remains relatively unaffected by the bulk mass and is given by
$m_{\Phi_n} = (n+\frac{\alpha_\Phi}{2} -\frac{3}{4}) \pi~k~e^{-\pi kR}$, 
with $\alpha_\Phi = \sqrt{4+\frac{\Lambda^2}{k^2}} \simeq 2 + 
\frac{\Lambda^2}{4 k^2}$. 
This can also be understood as follows:  Adding a bulk mass term $\Lambda$ 
has the effect of locally adding a mass term $\Lambda e^{-k y}$ at any point 
$y$ in the bulk. Since the zero mode of a massless bulk scalar field is 
localised towards $y\simeq 0$ the affect of adding a bulk mass shifts the
mass of the zero mode by $\sim \Lambda$. On the other hand the Kaluza-Klein 
modes are localised near $y=\pi R$, so adding the bulk mass term
affects them by an amount $\sim\Lambda e^{-\pi kR}$. Thus the Kaluza-Klein 
spectrum of the $\phi$ and the PV field is approximately the same while 
zero modes are separated by a mass scale $\Lambda$.

The corrections to the photon self energy $\Pi_{\mu\nu}(q^2)= (q^2 
\eta_{\mu\nu} - q_\mu q_\nu)\Pi(q^2)$ are given by~\cite{pomarol}
\begin{eqnarray}
      \Pi(0) &\propto& \sum_n \int d^4p~\left[\frac{1}{(p^2+m_{\phi_n}^2)^2}
    -\frac{1}{(p^2+m_{\Phi_n}^2)^2}\right]~,\nonumber\\
    &\simeq& \frac{b_\phi}{8\pi^2}\log\frac{\mu}{\Lambda} -
      \frac{b_\phi}{64\pi^2} \frac{\Lambda^2}{k^2}(\pi k R)~,
\end{eqnarray}
where we have introduced an infrared cutoff $\mu$ and $b_\phi=1/3$ is the
$\beta$-function coefficient. For $\mu\ll \Lambda < k$
the Kaluza-Klein contribution is negligible since the contributions
from the Kaluza-Klein modes effectively cancels out. Instead the dominant 
contribution arises purely from the zero modes and is given by
\begin{equation}
\label{logrun}
      \Pi(0) \simeq \frac{b_\phi}{8\pi^2}\log\frac{\mu}{\Lambda}~.
\end{equation}
This logarithmic dependence on the cutoff $\Lambda$ is exactly what we would 
obtain in four dimensions and consistently agrees with the AdS/CFT 
correspondence.

This result can be contrasted with what one obtains in 5D flat space. If
we add a bulk mass $\Lambda$ to a bulk scalar field the whole Kaluza-Klein 
tower shifts by an amount $\sim\Lambda$. Only the Kaluza-Klein 
states above this scale cancel with the PV field, leaving behind the 
contribution from the Kaluza-Klein states below $\Lambda$. This leads to
the contribution
\begin{equation}
      \Pi(0) \simeq \frac{b_\phi}{8\pi^2}\Lambda R~,
\end{equation}
which are the power-law corrections~\cite{ddg}.

It is also important to note that (\ref{logrun}) can only be understood
as the running of gauge couplings if matter is localised towards the UV 
brane. In this case the effective (field theory) 
cutoff is the Planck scale and the 
couplings can be interpreted as a running up to that 
scale. However for matter localised on the IR brane the
effective (field theory) cutoff is the TeV scale. Above this scale 
the effective field theory description breaks down and there
will be contributions from the fundamental (string) 
theory~\cite{pomarol,gr}.

\subsection{Partly Supersymmetric Grand Unification}

Let us now consider gauge coupling unification in the partly
supersymmetric standard model. The dominant contribution
to the running will be logarithmic and it is just a question of 
determining the $\beta$-function coefficients. Consider a 5D SU(5) 
gauge theory that is broken on the Planck brane, but preserved on 
the IR brane. We can implement this setup by imposing boundary
conditions on the bulk fields. These can be either Neumann $(+)$ or 
Dirichlet $(-)$ conditions corresponding to even or odd reflections, 
respectively, about the orbifold fixed points. 

The SU(5) gauge bosons form an ${\cal N}=2$ vector multiplet, 
${\cal V}=(V,S)$, where $V(S)$ is an ${\cal N}=1$ vector (chiral) 
multiplet. These fields are assumed to have the boundary conditions
\begin{equation}
  V=\left[\begin{tabular}{c}
            $V_\mu^a(+,+)$\\ 
            $V_\mu^A(-,+)$
            \end{tabular}\right]~,\qquad\qquad
  S=\left[\begin{tabular}{c}
            $S^a(-,-)$\\ 
            $S^A(+,-)$
            \end{tabular}\right]~,
\end{equation}
where the indices $a (A)$ run over the unbroken (broken) generators, 
and the first (second) argument refers to the UV (IR) boundary condition.
These boundary conditions break the SU(5) symmetry on the UV brane
and the only zero modes are the Standard Model gauge bosons $V_\mu^a(+,+)$.
At the nonzero mode level, the $X,Y$ gauge bosons (contained in 
$V_\mu^A(-,+)$), and the SU(5) adjoint scalar states all obtain TeV-scale 
masses.

Similarly, the Higgs sector is supersymmetric and contains two Higgs 
doublets which are embedded into two ${\cal N}=2$ bulk hypermultiplets, 
${\cal H}=(H_5,H^c_5)$ and $\bar {\cal H}=(\bar H_5,
\bar H^c_5)$, each transforming in the ${\bf 5}$ of SU(5). 
The boundary conditions are
\begin{equation}
  H_5 =\left[\begin{tabular}{c}
            $H_2(+,+)$\\ 
            $H_3(-,+)$
            \end{tabular}\right]~, \qquad\qquad
  H^c_5 =\left[\begin{tabular}{c}
            $H_2^c(-,-)$\\ 
            $H_3^c(+,-)$
            \end{tabular}\right]~,
\end{equation}
and similarly for $\bar {\cal H}$. The only zero modes are the
two Higgs doublets, $H_2(+,+)$ and ${\bar H}_2(+,+)$, as in the MSSM.
This choice of boundary conditions neatly solves the doublet-triplet 
splitting problem~\cite{kawamura}.

The Standard Model matter fields are also embedded into bulk ${\cal N}=2$ 
hypermultiplets. However, while it would seem natural to put all the fermions 
of one generation into a single ${\bf 5}$ and ${\bf 10}$, parity 
assignments actually require the quarks and leptons to arise from different 
SU(5) bulk hypermultiplets~\cite{hn1,hmr,bhn}. Thus, for each 
generation we will suppose that there are bulk hypermultiplets
$({\bf 5}_1,{\bf 5}_1^c)+({\bf 5}_2,{\bf 5}_2^c)$ and
$({\bf 10}_1,{\bf 10}_1^c)+({\bf 10}_2,{\bf 10}_2^c)$ with boundary conditions
\begin{eqnarray}
  &&{\bf 5}_1 = L_1(+,+) + d_1^c(-,+)~,\quad
    {\bf 5}_1^c = L_1^c(-,-) + d_1(+,-)~, \\
  &&{\bf 5}_2 = L_2(-,+) + d_2^c(+,+)~,\quad
    {\bf 5}_2^c = L_2^c(+,-) + d_2(-,-)~,\\
  &&{\bf 10}_1 = Q_1(+,+) + u_1^c(-,+) +e_1^c(-,+)~,\\
  &&{\bf 10}_1^c = Q_1^c(-,-) + u_1(+,-) +e_1(+,-)~,\\
  &&{\bf 10}_2 = Q_2(-,+) + u_2^c(+,+) + e_2^c(+,+)~,\\
  &&{\bf 10}_2^c = Q_2^c(+,-) + u_2(-,-) + e_2(-,-)~,
\end{eqnarray}
where the Standard Model fermions are identified with the zero modes of
the fields with (+,+) boundary conditions. Notice that this embedding
elegantly explains why the fermions need not satisfy the SU(5) mass 
relations and although each Standard Model generation arises from 
different ${\bf 5} + {\bf 10}$ fields, the usual charge quantization and 
hypercharge assignments are still satisfied~\cite{hn1,hmr}. This feature 
also explains why tree-level proton decay is not a problem in these models. 
There is simply no allowed coupling between $X,Y(-,+)$ gauge bosons and 
Standard Model fields $L_1(+,+)$ and $d_2^c(+,+)$ that is even under the 
orbifold symmetry. This is also true for couplings between Standard Model 
particles and the coloured Higgs triplets. 
However, a bulk U(1) symmetry must be introduced in order to prevent proton 
decay from higher-dimensional operators~\cite{wgu,ads}.

The specific contributions to the gauge couplings are given by
\begin{equation}
\label{gceqn}
     \frac{1}{g_a^2(p)} = \frac{\pi R}{g_5^2} + \frac{1}{g_{Ba}^2(\Lambda)} 
     + \frac{1}{8\pi^2}\Delta_a(p,\Lambda)~,
\end{equation}
where $g_{Ba}$ are boundary couplings, 
and $\Delta_a$ are the one-loop corrections. 
The first term in (\ref{gceqn}) is the universal contribution from 
the tree-level gauge coupling $g_5$ in the bulk. The second term 
in (\ref{gceqn}) is an SU(5) violating term that follows from breaking 
the SU(5) symmetry on the Planck brane. It can be neglected because
$g_{Ba}(\Lambda)\simeq 4\pi$, since the theory is effectively strongly 
coupled at the scale $\Lambda \gappeq k$~\cite{bc,gns,gr2}.
Thus, we see that the dominant contributions to the gauge couplings
will arise from the logarithmically enhanced terms of $\Delta_a$. 
These terms cannot be obtained in the strongly coupled 4D dual theory,
but instead can be calculated using the bulk zero-mode Green 
functions~\cite{choi}. 

The exact Greens function expression can be expanded at low energies 
$p\lappeq$ TeV to obtain the dominant logarithmic contributions. For 
vector bosons the dominant term is 
\begin{equation}
\label{delV}
     \Delta^a({\cal V})=b^a_{\cal V}\ln \frac{k}{p}+\dots~,
\end{equation}
where $b^a_{\cal V} =(0, -\frac{22}{3}, -11)$.
Recall that $k\simeq M_P$ is the AdS curvature scale so that (\ref{delV})
is the usual logarithmic contribution from the Standard Model gauge bosons.
There are no corresponding gaugino zero mode contributions because these
modes have received a large supersymmetry breaking mass and decouple at low
energy.

Instead for the Higgs sector, the leading contribution for the 
Higgs doublet is
\begin{equation}
\label{delHpp}
     \Delta^a({\cal H}_{++})=T_a({\cal H}_{++})\ln \frac{T}{p}+\dots~,
\end{equation}
where $T_a({\cal H}_{++})=\left(\frac{3}{10},\frac{1}{2},0\right)$ and 
$T$ is shorthand for the TeV scale. In this case the leading contribution 
is again a logarithmn but it is small. This can be interpreted as being 
due to the fact that the Higgs doublet is a composite particle. On the other 
hand the Higgs triplet contributions for $m_{-+}\lappeq p \lappeq T$,
with $m_{-+}$ the lowest lying massive state, are
\begin{equation}
\label{delHmp}
     \Delta^a({\cal H}_{-+})=T_a({\cal H}_{-+}) \left[\frac{2}{3}
      \ln \frac{k}{p} +\ln \frac{T}{p}+\dots\right]~,
\end{equation}
where $T_a({\cal H}_{-+})=\left(\frac{1}{5},0,\frac{1}{2}\right)$.
There is now both a large and small logarithmic contribution.
The small contribution is from the composite states, while the large
contribution is due to elementary degrees of freedom which are required 
to form a Dirac state. These extra elementary states can also be inferred by 
directly studying the dual 4D theory~\cite{cp}. Thus, the total
Higgs contribution from both Higgs hypermultiplets $\cal H$ and 
$\bar{\cal H}$ is
\begin{equation}
\label{delH}
     \Delta^a({\cal H} +\bar{\cal H})=b^a_{{\cal H}+\bar{\cal H}}
       \ln \frac{k}{p} + \ln \frac{T}{p} + \dots~,
\end{equation}
where $b^a_{{\cal H}+\bar {\cal H}}=\left(\frac{4}{15},0,\frac{2}{3}\right)$.

Finally the first two matter generations are (predominantly) elementary 
and give rise to the contribution
\begin{equation}
\label{del12}
     \Delta^a({\bf 5}_i^{(I)}+{\bf 10}_i^{(I)})
      =\frac{4}{3} \ln \frac{k}{p}+\dots~.
\end{equation}
This is the usual universal contribution from one generation of Standard Model
fermions which form a complete SU(5) multiplet. On the other hand the third
generation is partly composite with composite states $t_R,b_R,\tau_R$ and
elementary states $t_L,b_L,\tau_L,{\nu_\tau}_L$. This gives the nonuniversal
contribution
\begin{equation}
\label{del3}
     \Delta^a({\bf 5}_i^{(3)}+{\bf 10}_i^{(3)})=b^a_{(3)}\ln \frac{k}{p}
     +\frac{4}{3}\ln \frac{T}{p}+\dots~,
\end{equation}
where $b^a_{(3)}=\left(\frac{8}{15}, \frac{8}{3}, \frac{4}{3}\right)$.
Clearly we see that the large logarithmic contribution that arises
from the elementary states introduces a differential running in the gauge 
couplings.

If we now add up all the $\Delta^a$ contributions (\ref{delV}), (\ref{delH}), 
(\ref{del12}) and (\ref{del3}) arising from the elementary states 
in the model, then at the leading log level we obtain the total 
contribution~\cite{psgu}
\begin{equation}
     \Delta^a = b^a_{\rm total} \ln\frac{k}{p} + \dots~,
\end{equation}
where $b^a_{\rm total}=\left(\frac{52}{15},-2,-\frac{19}{3}\right)$.
These $\beta$-function coefficient values give $B=0.793$, which allowing
for an approximately $10\%$ theoretical uncertainty, agrees with the 
experimental value (\ref{Bexpt}). Interestingly the partly composite third 
generation has 
restored the gauge coupling unification without the gauginos and Higgsinos.
\\
\\
$\bullet$ {\sc Exercise:} {\it Use the one-loop gauge coupling 
expressions in Ref~\cite{choi} for $p\lappeq$ TeV, to verify the 
individual logarithmic contributions in $\Delta^a$.}
\\
\\
Note that even though the model is partly supersymmetric, supersymmetry 
plays no role in obtaining gauge coupling unification because all 
the differential running contributions come from the UV-brane localised 
elementary sector which is nonsupersymmetric. This means that a similar 
mechanism will also work for an inherently nonsupersymmetric model such as 
the minimal composite Higgs model~\cite{acs}.

\section{Conclusion}\label{conclusion}
Warped models in a slice of AdS$_5$ provide a new framework to study 
solutions of the hierarchy problem at the TeV scale. The warp factor
naturally generates hierarchies and can be used to either stabilise the 
electroweak scale or explain why the scale of supersymmetry breaking is 
low. Remarkably by the AdS/CFT correspondence these 5D warped models are 
dual to strongly coupled 4D theories. The Higgs localised on the IR brane 
is dual to a composite Higgs. The corresponding Higgs boson mass can be light
compared to the IR cutoff by using either a global symmetry  and treating 
the Higgs as a pseudo Nambu-Goldstone boson or using supersymmetry to make
only the Higgs sector supersymmetric. The good news is that these models 
are testable at the LHC (and an eventual linear collider), so it will be 
an exciting time to discover whether 
Nature makes use of the fifth dimension in this novel way. If not, there is 
no bad news, because the warped fifth dimension literally provides a new 
theoretical framework for studying the dynamics of strongly coupled 4D gauge 
theories and this will be an invaluable tool for many years to come.

\section*{Acknowledgements}

I would like to thank the organisers for organising such an excellent summer
school. I am especially grateful to Alex Pomarol for collaboration on the 
topics which formed the basis of these lectures. I also thank Kaustubh Agashe,
Roberto Contino, Yasunori Nomura, Erich Poppitz and Raman Sundrum for  
discussions on warped matters, and Brian Batell for comments
on the manuscript. This work was supported in part by a Department of Energy 
grant DE-FG02-94ER40823, a grant from the Office of the Dean of the 
Graduate School of the University of Minnesota, and an award from Research 
Corporation.

%
\end{document}